\documentclass[12pt]{iopart}
\pdfoutput=1

\expandafter\let\csname equation*\endcsname\relax
\expandafter\let\csname endequation*\endcsname\relax

\usepackage{amsmath,amsfonts,amssymb}
\usepackage{graphicx}

\newcommand{\be}{\begin{equation}}
\newcommand{\ee}{\end{equation}}
\newcommand{\bea}{\begin{eqnarray}}
\newcommand{\eea}{\end{eqnarray}}
\def\neff{N_{\rm eff}}
\def\dN{\Delta N_{\rm eff}}
\def\paenpe{PA{\scriptsize RTH}EN{\scriptsize O}PE}
\def\nn{\nonumber}

\begin{document}
\setlength{\unitlength}{1mm}

\title[Primordial Deuterium after LUNA: concordances and error budget]{Primordial Deuterium after LUNA: concordances and error budget}

\author{Ofelia Pisanti, Gianpiero Mangano, Gennaro Miele, and Pierpaolo Mazzella}

\address{Dipartimento di Fisica E. Pancini, Universit\`a di Napoli Federico II, and INFN, Sezione di Napoli, Via Cintia, I-80126 Napoli, Italy}


\begin{abstract}

The accurate evaluation of the nuclear reaction rates and corresponding uncertainties is an essential requisite for a precise determination of light nuclide primordial abundances. 
The recent measurement of the \mbox{D(p, $\gamma)^3$He} radiative capture cross section by the LUNA collaboration, with its order 3\% error, represents an important step in improving the theoretical prediction for Deuterium produced in the early universe. In view of this recent result, we present in this paper a full analysis of its abundance, which includes a new critical study of the impact of the other two main processes for Deuterium burning, namely the deuteron-deuteron transfer reactions, D(d, p)$^3$H and D(d, n)$^3$He. In particular, emphasis is given to the statistical method of analysis of experimental data, to a quantitative study of the theoretical uncertainties, and a comparison with similar studies presented in the recent literature. We then discuss the impact of our study on the concordance of the primordial nucleosynthesis stage with the Planck experiment results on the baryon density $\Omega_bh^2$ and the effective number of neutrino parameter $\neff$, as function of the assumed value of the $^4$He mass fraction $Y_p$. While after the LUNA results, the value of Deuterium is quite precisely fixed, and points to a value of the baryon density in excellent agreement with the Planck result, a combined analysis also including Helium leads to two possible scenarios with different predictions for $\Omega_bh^2$ and $\neff$. We argue that new experimental results on the systematics and the determination of $Y_p$ would be of great importance in assessing the overall concordance of the standard cosmological model.

\end{abstract}


\maketitle


\section{Introduction}

The (present) baryonic energy fraction $\omega_b\equiv \Omega_bh^2$ is measured with remarkable precision thanks to the observations of the Planck experiment on Cosmic Microwave Background radiation (CMB) \cite{Planck2018},
\bea
\omega_b &=& 0.02237\pm 0.00015 \,\,\, {\rm Planck}, \label{Planck} \\ 
\omega_b &=& 0.02242\pm 0.00014 \,\,\, {\rm Planck + BAO} \label{Planck+BAO}.
\eea
Actually, the first determination of the baryon content of the universe was obtained using Big Bang Nucleosynthesis (BBN) decades ago. Assuming that there are no changes in the baryon to photon density ratio from BBN down to CMB formation epoch, the precise CMB determination of $\omega_b$ implies that nowadays the theory of BBN, in its minimal scenario, is a parameter free model and, as such, has become a high precision tool to investigate effects in the early universe of physics beyond the Standard Model and/or exotic cosmological models (see for example \cite{Iocco:2008va,Pitrou:2018cgg}).

With the exception of $^7$Li, for which predictions are a factor $\sim 3$ higher than observations, there is a good agreement between the predicted and observed primordial abundances of the light elements, an agreement which should be quantified and updated in view of new experimental results and theoretical insights. The astrophysical determination of Deuterium has now reached a percent accuracy, D/H=$(2.527 \pm 0.030)\times 10^{-5}$ at 68\% of C.L. \cite{Cooke:2017cwo}, and $^4$He uncertainty has been reduced to less than 2\%, $Y_p= 0.2446\pm 0.0029$ \cite{Peimbert:2016bdg}, $Y_p= 0.2449\pm 0.0040$ \cite{Aver:2015iza}, $Y_p= 0.2551\pm 0.0022$ \cite{Izotov:2014fga}. The most recent result is $Y_p= 0.2436 \pm 0.0040$ \cite{Hsyu:2020uqb}, very close to the results of  \cite{Peimbert:2016bdg,Aver:2015iza}. However, while the $^4$He abundance prediction is limited by  systematics effects in astrophysical observations -- notice also the discrepancy between the estimates of \cite{Izotov:2014fga} and \cite{Peimbert:2016bdg,Aver:2015iza,Hsyu:2020uqb} -- Deuterium measurements are instead expected to increase their precision by about an order of magnitude \cite{Cooke:2016rky}, using 30 m class telescopes. 

One of the most important issues to address in order to get an accurate determination of light nuclide abundances and their  corresponding uncertainties is the evaluation of the nuclear reaction rates entering the BBN reaction network. The estimate of the uncertainties on the primordial light elements yields comes from a quite involved error propagation process of the several BBN input parameters. All most recent BBN analysis (see for example \cite{DiValentino:2014cta,Marcucci:2015yla,Pitrou:2018cgg,Fields:2019pfx}) agree that the largest contribution to the error on primordial Deuterium comes from the uncertainty on three main destruction process rates: the D(p,$\gamma)^3$He radiative capture and the Deuteron-Deuteron transfer reactions, D(d, n)$^3$He and D(d, p)$^3$H. We will refer to these rates per unit density of incoming particles as $R_{dp\gamma}$, $R_{ddn}$, and $R_{ddp}$, respectively (with $R_{pn\gamma}$ denoting the p(n,$\gamma$)$^2$H rate). On the other hand, the value of $Y_p$ is almost entirely fixed by the value of the neutron lifetime $\tau_n$ since this parameter fixes the freezing time of the neutron-proton chemical equilibrium and thus, the relic abundances of neutrons, which eventually get bound in $^4$He nuclei. In the following, we will adopt the most recent average and error for its value $\tau_n = (879.4 \pm 0.6)$ \cite{Zyla:2020zbs}, which translates into a very small theoretical error on $ \Delta Y_p \sim \pm 0.00012$, much smaller than the experimental uncertainty on $Y_p$ from astrophysical measurements, and thus, completely negligible. 
 
A very recent breakthrough in BBN has been the new underground measurement of the D(p,$\gamma)^3$He cross section by the LUNA Collaboration \cite{Cavanna:2019pme} at the Gran Sasso National Laboratory in Italy, which reached a precision of the order of 3\% in the center of mass energy relevant for the BBN dynamics, and which allowed to have an estimate of $R_{dp\gamma}$ with unprecedented precision \cite{Mossa:2020qgj,Nature}. Since this rate is one of the leading parameter in quantifying the amount of Deuterium burning, the smaller uncertainty on $R_{dp\gamma}$ is now fixing the D/H ratio for a fixed baryon density, with a decreased error, a factor two larger than the one of its astrophysical measurement. Furthermore, using the LUNA measurement, which dominates a global fit of other available data, leads to a value of the S-factor which is sensibly higher than what was obtained by previous analysis of available data \cite{Adelberger:2010qa}. As discussed in \cite{Nature}, using the result of \cite{Cooke:2017cwo} allows for a determination of the baryon density $\omega_b$ using the D/H abundance {\it only} which is in excellent agreement with the Planck results and {\it with a comparable uncertainty}.

In this paper we perform a new analysis of the main leading reactions responsible for Deuterium burning during BBN. We address first the new LUNA results on BBN. This issue has been already discussed in details in \cite{Nature}, but we report it here for the sake of completeness and also to include further technical details. In particular, we describe in details our method of statistical analysis and data selection criterium. We also present a new estimation of the ddn and ddp rates, which have also been used in the BBN study of \cite{Nature}. Another aim of this paper is to compare our method and results with other similar analysis, trying to asses a "concordance" benchmark estimate of theoretical uncertainties, which desirably should account for different statistical treatment of data and different dataset selection. This can be viewed as a contribution in the direction of reaching a shared trustable theoretical determination of primordial Deuterium, which might be used as a robust  cosmological probe for future analyses by different groups.

The paper is structured as follows. In Section \ref{statistics} we describe the method used to deduce $R_{dp\gamma}$, $R_{ddn}$, and $R_{ddp}$ from experimental datasets and the data selection, while in Section \ref{sfactor} we present our results, and compare with other approaches reported in the recent literature. In Section \ref{paenpe} we discuss the consequences of our study on Deuterium theoretical value and its uncertainty, obtained by an updated version \cite{Gariazzo:2019} of the {\paenpe} public code \cite{Pisanti:2007hk,Consiglio:2017pot} which now includes all changes on rate values and uncertainties discussed here. Section \ref{cosmology} describes the cosmological implications of our Deuterium theoretical determination and the impact of different $^4$He measurement. We give our conclusions in Section \ref{conclusions}.

\section{Nuclear network: analysis method and data selection}
\label{statistics}

The thermal average of the cross section times relative velocity for a two body $i+j$ non resonant reaction induced by a charged particle (the ones we are mainly interested here) is \cite{Fowler:1967ty}
\be
\left \langle \sigma v \right \rangle(T)=\sqrt{\frac{8}{\pi\mu_{ij}}}T^{-3/2}\int_{0}^{+\infty}\mathrm{d} E\,E \,\sigma(E) e^{-E/T},
\label{sigmav}
\ee
where $\mu_{ij}$ is the reduced mass of the $i+j$ system and $E$ is the kinetic energy in the center of mass (CM) frame. The previous equation is usually cast in a more convenient form, extracting from the cross section $\sigma(E)$ the exponential suppression of the penetrability (Gamow) factor $e^{-\sqrt{E_G/E}}$, where the Gamow energy is $E_G\equiv 2\pi^2 \mu_{ij}(Z_iZ_j\alpha)^2$. The averaged cross section is then written in terms of the astrophysical S-factor, $S(E)$,
\be
S(E)\equiv\sigma(E) E\,e^{\sqrt{E_G/E}},
\ee
which represents the intrinsic nuclear part of the reaction probability. From the value of the onset energy of BBN and taking into account the energy dependence of the integral (\ref{sigmav}) one can conclude that the BBN relevant range of the CM kinetic energy is from $\sim 10$ keV to $\sim 400$ keV (see Section \ref{sensitivity} for more details on this topic).

For every single reaction the S-factor, in absence of a specific theoretical model, has to be fitted from data collected in different nuclear physics experiments, often covering only limited energy ranges which are partially, or entirely, not overlapping. The main difficulty is that these experiments have typically different normalization errors, due to different detection efficiency, ion beam current measurement, target thickness, etc., which in some cases are not even estimated. In the absence of well-defined theoretical prescriptions on this matter, different approaches have been used in the literature to determine the S-factors which  lead to slightly different outcomes.

In ref. \cite{DAgostini:1993arp} it is argued that, when data sets are affected by different normalization errors, the use of the covariance matrix approach in $\chi^2$ minimization produces a negative bias that systematically underestimates the physical quantities, because data sets with smaller statistical uncertainties dominates the $\chi^2$ even if they are affected by larger systematics. Consider the simple case of two physical quantities, $X_1$ and $X_2$, with measurements $x_i\pm \sigma_i$, affected by an error $\epsilon$ on a (unknown) normalization constant, $\omega$. Thus, one should consider the normalization factor as a further variable whose measurement would give $\omega=1$ in case of a well calibrated measure instrument, with uncertainty $\epsilon$. In the simplest case that the theoretical function to be fitted through the data points is a constant, $k$, using a standard $\chi^2$ minimization, equivalent to the covariance matrix formulation, we have
\be
\chi^2=\frac{(\omega x_1-k)^2}{\sigma_1^2} + \frac{(\omega x_2-k)^2}{\sigma_2^2} +\frac{(\omega -1)^2}{\epsilon^2},
\label{chi2dagostiniNO}
\ee
where data are rescaled independently of their errors. However, the expected value of $\omega$ is shown to be \cite{DAgostini:1993arp}
\be
\left \langle \omega\right\rangle = \frac{1}{1+(n-1)\,\epsilon},
\ee
with $n$ the number of data. The previous result shows an evident bias toward small values of $\left \langle \omega\right\rangle$, which increases when a larger number of data points is considered and depends on the size of the error $\epsilon$ on the scale factor. The unbiased result with $\widehat{k}=\overline{x}$, suggested in \cite{DAgostini:1993arp}, can be obtained with a similar $\chi^2$ form, but rescaling the errors too,
\be
\chi^2=\frac{(\omega x_1-k)^2}{(\omega \sigma_1)^2} + \frac{(\omega x_2-k)^2}{(\omega \sigma_2)^2} +\frac{(\omega -1)^2}{\epsilon^2}.
\label{chi2dagostini}
\ee

The $\chi^2$ implemented in the present work, as originally in \cite{Serpico:2004gx}, is the most natural generalization of Eq. (\ref{chi2dagostini}),
\be
\chi^2(a_l,\omega_k)=\sum_{i_k}\frac{(S_{th}(E_{i_k},a_l)-\omega_k\,S_{i_k})^2}{\omega_k^2\,\sigma_{i_k}^2}+\sum_k\frac{(\omega_k-1)^2}{\epsilon_k^2} \equiv \chi^2_{stat} + \chi^2_{norm}.
\label{chi2tot}
\ee
In the previous equation the (unknown) parameters found by the $\chi^2$ minimization are the coefficients $a_l$ in a polynomial expansion of the theoretical  S-factor, $S_{th}(E_{i_k},a_l)$, and the multiplicative normalization constants (one for each experiment), $\omega_k$. Moreover, $E_{i_k}$, $S_{i_k}$, $\sigma_{i_k}$, and $\epsilon_{k}$ are the (center of mass) energy, experimental S-factor data, their statistical error, and (relative) normalization uncertainty of the $i$-th data point of the $k$-th data set, respectively. Note that, if only a total error $\sigma_{i_k}^{tot}$ is available for the data points of a certain experiment, that error is used instead of $\sigma_{i_k}$, and the normalization uncertainty, $\epsilon_{k}$, is estimated as max[$\sigma_{i_k}^{tot}/S_{i_k}$].

The use of a polynomial expansion for the S-factor is justified by the fact that, apart from resonance contributions, it is a smooth function of energy. The sensitivity to the fitting form, i.e. the polynomial degree of $S$, is checked by increasing it until the minimum of the reduced $\chi^2$ stabilizes and typically, polynomials of 2-th or 3-th degree are a satisfying choice. The second term, $\chi^2_{norm}$, links the data normalization factor $\omega_k$ (the "nuisance" parameters) for each data set to the experimental estimate of the corresponding uncertainty, $\epsilon_k$. Note that all experimental points, $S_{i_k}$, of the same data sets are correlated by sharing the same normalization, $\omega_k$, and that the contribution of the penalty factor, $\chi^2_{norm}$, does not allow $|\omega_k-1|$ to be much greater than the estimated or quoted normalization errors, $\epsilon_k$.

A polynomial fit to the data is employed also by the authors of \cite{Fields:2019pfx}, which however show also the impact on the results of adopting a theory-based rate for the D(p,$\gamma)^3$He reaction. On the other hand the authors in \cite{Coc:2015bhi}\footnote{Note that the more recent results of \cite{Pitrou:2018cgg}, which take advantage of the use of Bayesian statistics of \cite{Inesta:2017jhc}, turn out to be in agreement with those of \cite{Coc:2015bhi} at 1\% level in the temperature range of BBN.} use, instead, the following $\chi^2$ expression,
\be
\chi^2 (\alpha_k)=\sum_{i_k}\frac{[S_{i_k}-\alpha_kS_{th}(E_{i_k})]^2}{\sigma_{i_k}^2},
\label{chi2pitrou}
\ee
where the fitting functions, $S_{th} (E)$, come from nuclear reaction models, and the single free parameters of the fit are the normalization constants, $\alpha_k$, one for each data set. Then, the global normalization for a given reaction measured in multiple data sets is given by the weighted average of the $\alpha_k$. Notice that, apart for the presence of the $\chi^2_{norm}$ term, Eq. (\ref{chi2tot}) is equivalent to Eq. (\ref{chi2pitrou}) for $\alpha_k=1/\omega_k$. A similar procedure for calculating the global normalization of data has been used in \cite{Cyburt:2015mya}, where the fitting functions, $S_{th} (E)$, are the NACRE-II \cite{Xu:2013fha} theoretical S-factors.

A recent approach to the analysis of nuclear rates is based on Bayesian probability theory, which directly gives the probability density functions for the reaction rates (see \cite{Inesta:2017jhc,Iliadis:2016vkw} for the reactions we are interested in). The results seem to be in fair agreement with the frequentist analyses for both the recommended values of the S-factors and the magnitude of the uncertainties.

Besides the statistical analysis method, Eq.s (\ref{chi2tot}) and (\ref{chi2pitrou}), a possible source of disagreement among different authors may lie in the data selection. In particular, in \cite{Coc:2015bhi} a strict data selection was made, excluding from the fit all experiments for which the systematic uncertainty was not quoted or too large. A further data selection was made to exclude all data points falling outside the energy validity range of the theoretical model or with energy dependence different from the theoretical expectation.

While the use of a theoretical prejudice is in principle helpful to point out possible systematics effects in experimental data, on the other hand excluding points with larger errors at the aim of reducing uncertainties in parameter extraction is not a good strategy,  also in view of the fact that one expects that the more precise data will dominate the $\chi^2$ minimization anyway. We therefore, chose to include in our analysis all data on deuteron-deuteron transfer reactions, including those obtained via the Trojan Horse (TH) method \cite{Tumino:2014}, which had been excluded by the authors of \cite{Coc:2015bhi},  who claim that they do not comply with the theoretical expectation of \cite{Arai:2011zz}. Actually, as we will see in the following sections, when uncertainties are taken into account there is no compelling evidence against the use of TH data.

Of course we also include in our analysis the new data  \cite{Mossa:2020qgj,Nature} on \mbox{D(p, $\gamma)^3$He} radiative capture from LUNA experiment, as well as those of ref. \cite{Tisma:2019acf}.

\section{Astrophysical S-factors and rates}
\label{sfactor}

In this section we discuss the results on astrophysical S-factors and thermal rates for the Deuterium burning reactions.

After $\chi^2$ minimization using the CERN library MINUIT package \cite{minuit}, we obtain the best fit values for the coefficients, $\widehat{a}_l$ of the $S_{th}(E)$ polynomial functions, as well as their covariance matrix, cov$(a_i,a_j)$. The average cross sections times velocity $\langle \sigma v \rangle(T) \equiv R(T)$, i.e. the thermal rates per unit density of incoming particles, are then calculated by numerical integration of $S_{th}(E)$ convoluted with the appropriate Boltzmann/Gamow kernel $K(E,T)$ (see Eq.\,(\ref{sigmav})),
\be
R(T)=\int_{0}^{\infty}\mathrm{d} E\,K(E,T)\,S_{th}(E,\widehat{a}),
\label{rate}
\ee
and the (squared) rate error $\Delta R^2$ through the standard error propagation as
\be
\Delta R^2(T)=\int_{0}^{\infty}\mathrm{d} E'\,K(E',T)\int_{0}^{\infty}\mathrm{d} E\,K(E,T)\,\sum_{i,j}\frac{\partial S_{th}(E',a)}{\partial a_i} \bigg|_{\widehat{a}}\frac{\partial S_{th}(E,a)}{\partial a_j} \bigg|_{\widehat{a}} \text{cov}(a_i,a_j).
\label{rateerror}
\ee
We then derive the total uncertainty band on the thermal rates, as described in detail in \cite{Serpico:2004gx}, by inflating the estimated error by the usual factor $\sqrt{\chi^2}$ and summing to it in quadrature the overall scale error for the given reaction.

We discuss in the following the results for the three processes that we have re-analyzed, while for the neutron-proton fusion reaction \mbox{p(n, $\gamma)^2$H}, for which a very accurate result has been derived using pionless effective field theory \cite{Rupak:1999rk}, we use the same value discussed in \cite{Serpico:2004gx}. 

\subsection{Reaction D(p,$\gamma)^3$He}

The new analysis of the D(p,$\gamma)^3$He has been already presented in \cite{Nature}. We include it here for completeness and to discuss  details which for brevity were not covered in \cite{Nature}.

The radiative capture D(p,$\gamma)^3$He, from now on dp$\gamma$, is a fundamental process of $^3$He synthesis in many astrophysical contexts, like nuclear fusion in stars as one of the steps of the p-p chain, and is also a leading reaction in BBN for Deuterium burning and Helium production.

The astrophysical S-factor at low energy, at the solar Gamow peak $E_{G}\simeq 9$ keV, has been measured with a remarkable precision \cite{Casella:2002yej}, but for the BBN energy range (10-400 keV) the experimental situation was unclear, because of the paucity of the available experimental data \cite{Ma:1997zza} and their disagreement with the phenomenological fit of $S(E)$ for $E\simeq 0-2$ MeV \cite{Adelberger:2010qa}, from now on AD2011. Its uncertainty was about $6\%\div10\%$ in the energy range relevant for BBN (see \cite{DiValentino:2014cta} for a detailed discussion) due to the lack of  data in this range, resulting in an error on the primordial D/H ratio of the order of 2.6\%. Moreover, using the AD2011 fit to compute the $D(p,\gamma)^3$He thermal rate leads to a value for Deuterium abundance higher than the experimental value. Presently, the understanding of this crucial cross section in the relevant energy range highly improved thanks to the LUNA experiment results at the underground Gran Sasso Laboratory \cite{Nature}.
\begin{figure}[t]
\begin{center}
\includegraphics[width=.6\textwidth]{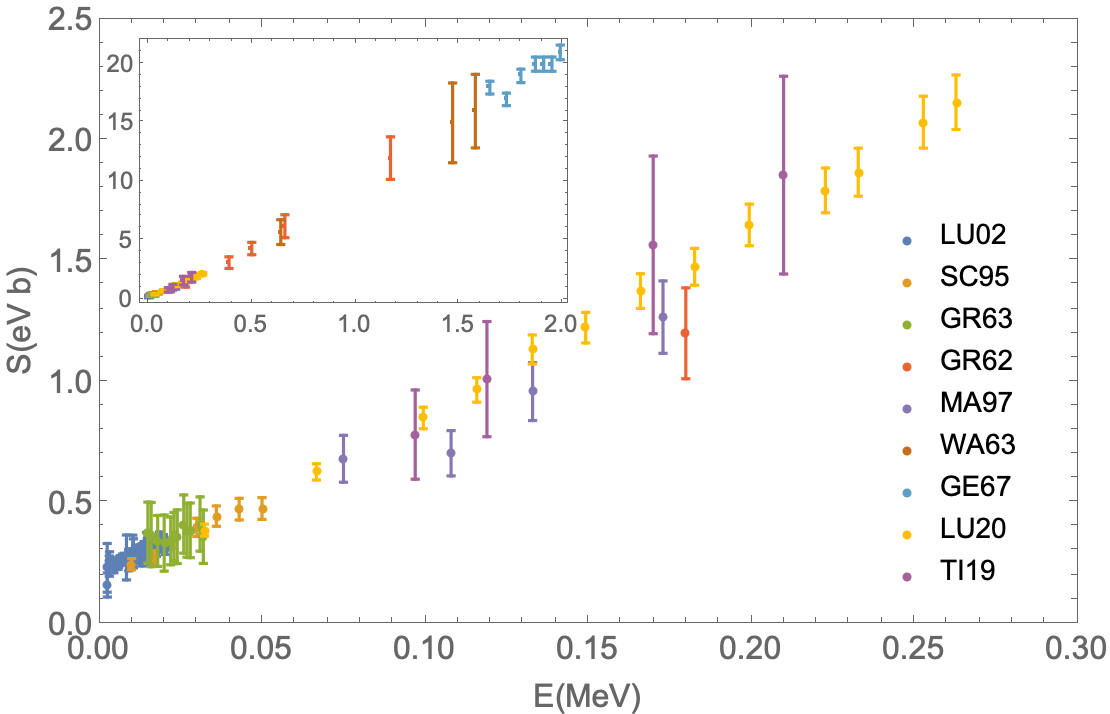}\hspace{2pc}
\end{center}
\begin{minipage}[b]{\textwidth}\caption{\label{dpgamma_data}Data sets on dp$\gamma$ S-factor versus center of mass energy in the BBN relevant range. The inset shows higher energy data.}
\end{minipage}
\end{figure}

It was conjectured, before LUNA results, that a way to get a better agreement between Deuterium abundance theoretical prediction and its measured value is of course to increase the value of the dp$\gamma$ astrophysical S-factor \cite{Nollett:2011aa, DiValentino:2014cta}. This possibility was supported by the {\it ab initio} calculation of \cite{Marcucci:2015yla}, from now on MAR2015. In this approach, a nuclear Hamiltonian is adopted retaining both two- and three-nucleon interactions, the Argonne v18 and the Urbana IX, respectively. MAR2015 rate is an upgrade of a previous determination \cite{Marcucci:2005zc}, from now on MAR2005, where higher order terms had been added to the nuclear Hamiltonian as long as a first estimate of the theoretical uncertainty.  Further developments are expected in the near future. Note that MAR2015 rate is about 8\% larger than MAR2005 one, and about 15\% larger than the AD2011 fit. The theoretical behaviour of MAR2005 is adopted by the authors of \cite{Iliadis:2016vkw}, from now on IL2016, in a Bayesian analysis of selected experimental datasets resulting in S-factor and rate with the same normalization of the MAR2005 determination.

\begin{figure}[p]
\begin{center}
\includegraphics[width=.5\textwidth]{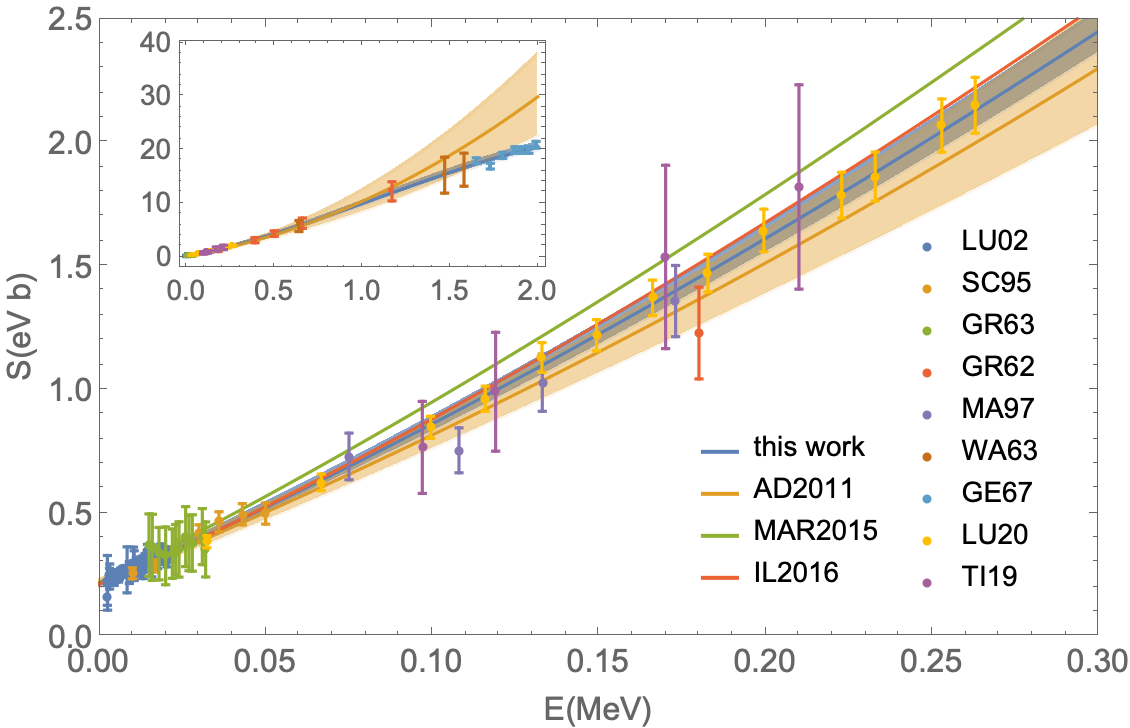}\hspace{2pc}
\end{center}
\begin{minipage}[b]{\textwidth}\caption{\label{S_dpgamma}(Color online) The dp$\gamma$ S-factor versus center of mass energy. Note that the experimental points have been renormalized by the corresponding $\omega_k$ obtained by the global fit. The curves are the data fit derived in this work (blue line and band), the fit of \cite{Adelberger:2010qa} (orange line and band), the {\it ab initio} theoretical calculation of \cite{Marcucci:2015yla} (green), and the determination of \cite{Iliadis:2016vkw} (light red). Bands are the 1-$\sigma$ uncertainty ranges.}
\end{minipage}
\end{figure}
\begin{figure}[p]
\begin{center}
\includegraphics[width=.5\textwidth]{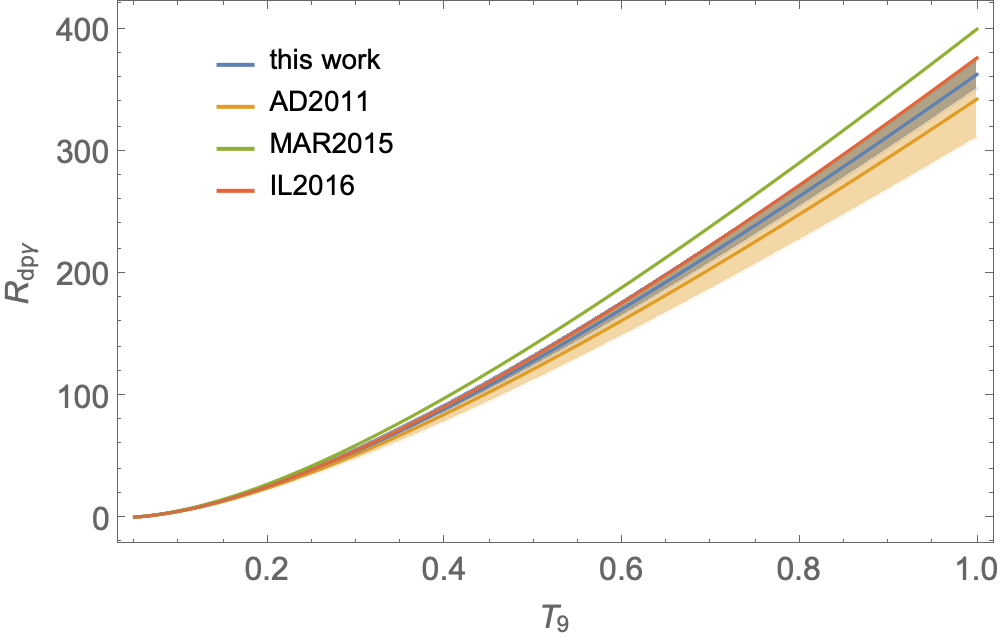}\hspace{2pc}
\end{center}
\begin{minipage}[b]{\textwidth}\caption{\label{R_dpgamma}(Color online) $R_{dp\gamma}$ rate versus the temperature in $10^9$ K. Conventions are the same of Fig. \ref{S_dpgamma}.}
\end{minipage}
\end{figure}
\begin{figure}[p]
\begin{center}
\includegraphics[width=.5\textwidth]{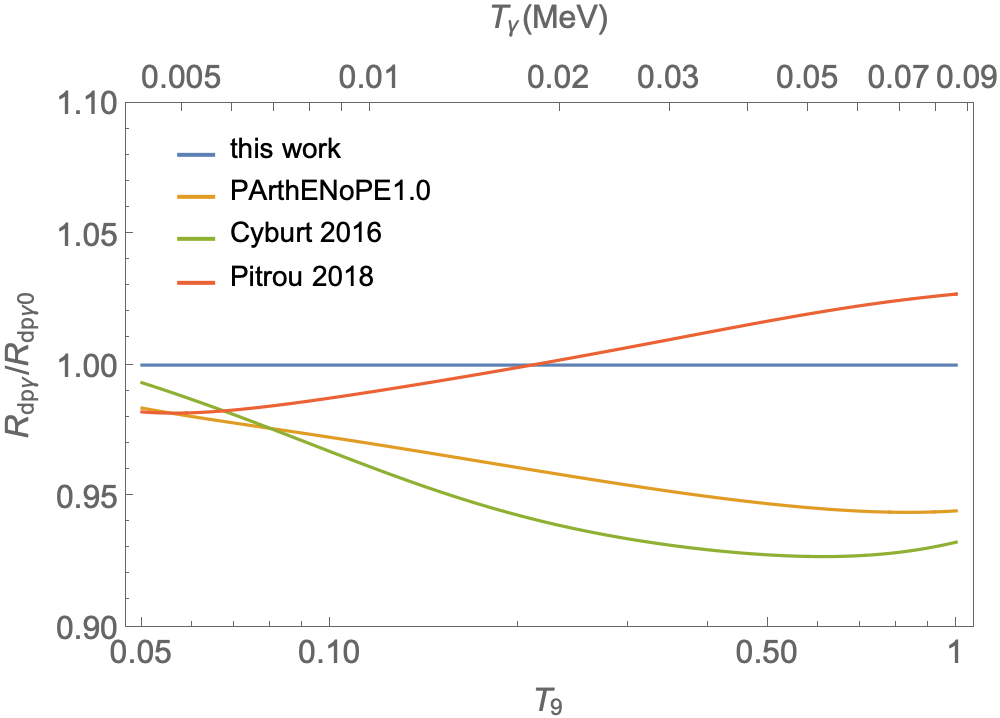}\hspace{2pc}
\end{center}
\begin{minipage}[b]{\textwidth}\caption{\label{dpgamma}Ratios between different $R_{dp\gamma}$ rates and its value obtained in this work, $R_{dp\gamma0}$, versus the temperature in $10^9$ K (see text).}
\end{minipage}
\end{figure}

We have re-analyzed all the data on this process \cite{Casella:2002yej,Ma:1997zza,Griffiths:1962,Griffiths:1963,Schmid:1997zz,Warren:1963zz,Geller:1967baf} (reported in Fig.\ref{dpgamma_data}) by adding to our previous selection the data of \cite{Tisma:2019acf} (with small impact on the result of the fit) and the new results of LUNA experiment \cite{Nature}. Fig. \ref{S_dpgamma} shows the result of the fit (blue line and blue 1-$\sigma$ uncertainty band) compared to the old determination AD2011 (orange line and orange 1-$\sigma$ uncertainty band), to the theoretical value MA2015 (green line), and to the result of IL2016 (light red line).  As it can be seen from the plot, in the considered energy range the fit is almost entirely dominated by LUNA results. The S-factor together with the normalization factors $\omega_k$ for each data set are reported in \ref{A1}. The corresponding value of $R_{dp\gamma}$ with its 1-$\sigma$ uncertainty as function of the temperature $T_9\equiv T/(10^9\,\text{K})$ is shown in Fig. \ref{R_dpgamma}. The results of our fitting procedure are compared to the other findings in Fig. \ref{dpgamma}, where we show the ratio with our present calculation of our previous analysis \cite{Pisanti:2007hk} (\paenpe{1.0}), of \cite{Cyburt:2015mya} (Cyburt 2016), and of \cite{Pitrou:2018cgg} (Pitrou 2018). Note that in \paenpe{1.0}, as in \cite{Cyburt:2015mya}, the AD2011 rate \cite{Adelberger:2010qa} was implemented, which amounts to a 7\% maximum difference with our present determination. A 3\% difference, instead, is found with the results of \cite{Pitrou:2018cgg}, which adopt the rate given by the S-factor of \cite{Iliadis:2016vkw} corresponding to the {\it ab initio} functional form of \cite{Marcucci:2005zc}.

\subsection{Reactions D(d,p)$^3$H and D(d,n)$^3$He}

\begin{figure}[b]
\begin{center}
\begin{minipage}{.45\textwidth}
\includegraphics[width=1.\textwidth]{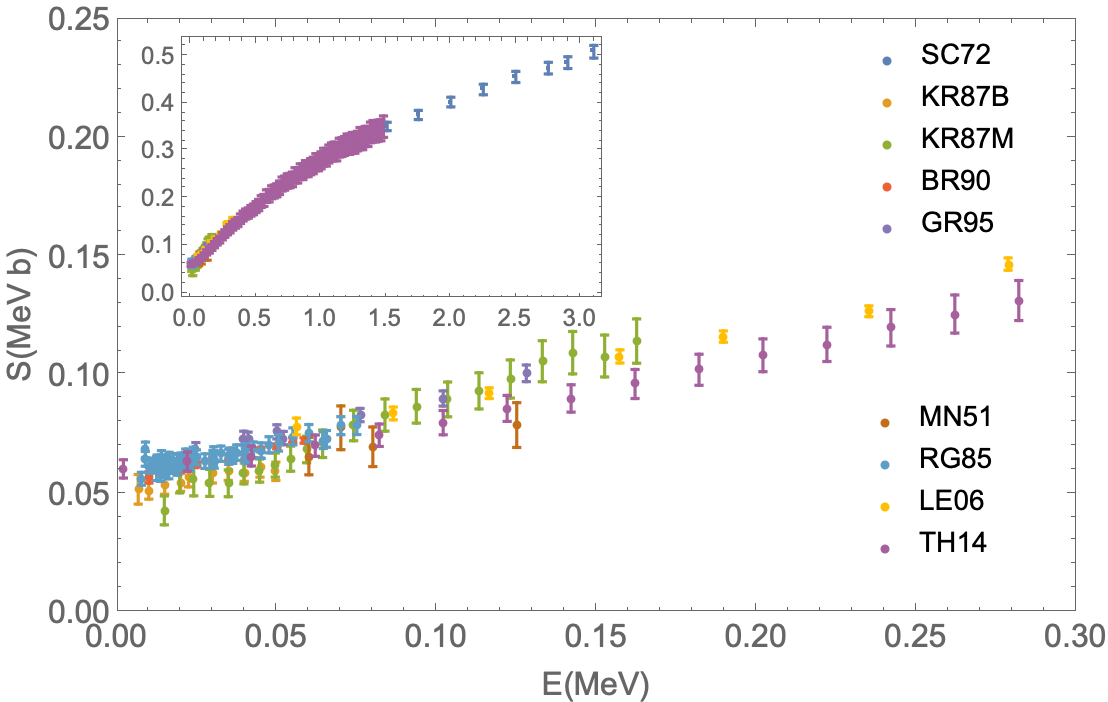}
\end{minipage}\hspace{2pc}%
\begin{minipage}{.45\textwidth}
\includegraphics[width=1.\textwidth]{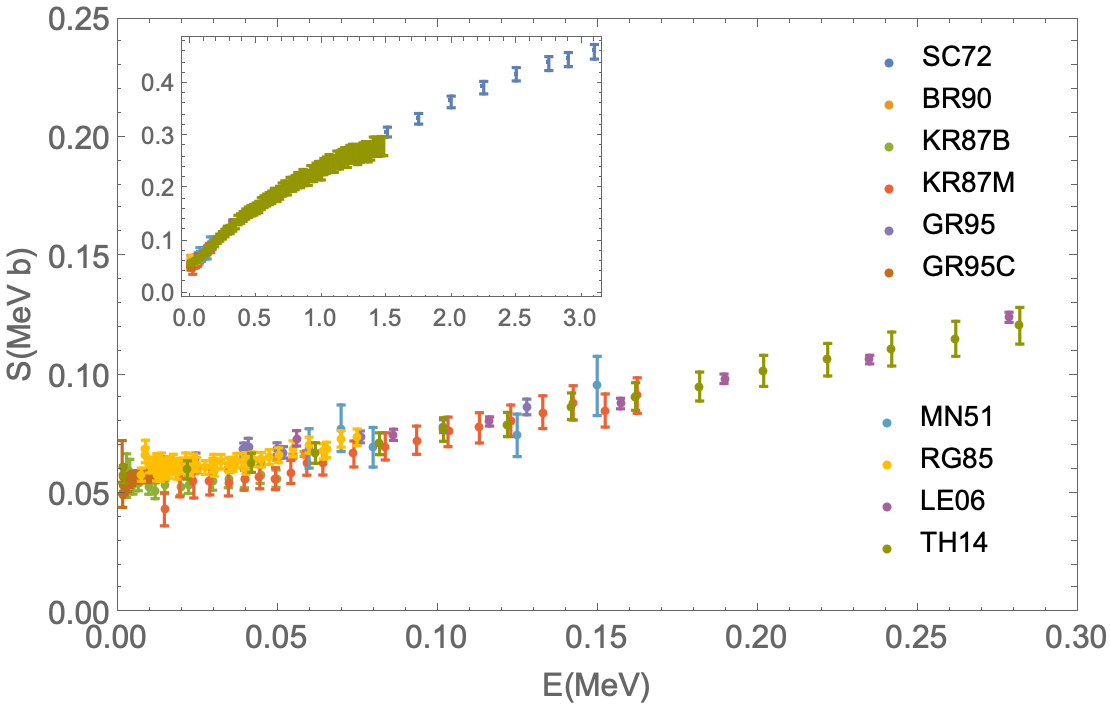}
\end{minipage} 
\end{center}
\caption{\label{dd_data}(Color online) Data sets on d-d transfer reactions S-factors versus center of mass energy in the BBN relevant range: ddn (left plot) and ddp (right plot). The inset shows higher energy data.}
\end{figure}
The D(d,p)$^3$H process (ddp in short) is the leading source of primordial tritium synthesis while at $T_9\sim 1$ almost all $^3$He is produced through the D(d,n)$^3$He channel\footnote{See \cite{Serpico:2004gx} for an analysis of the relevant nuclear reactions in the creation/destruction of a definite light nuclide.} (ddn in short). Both these reaction rates are strongly affecting the primordial Deuterium abundance and its theoretical uncertainty.

\begin{figure}[t]
\includegraphics[width=.5\textwidth]{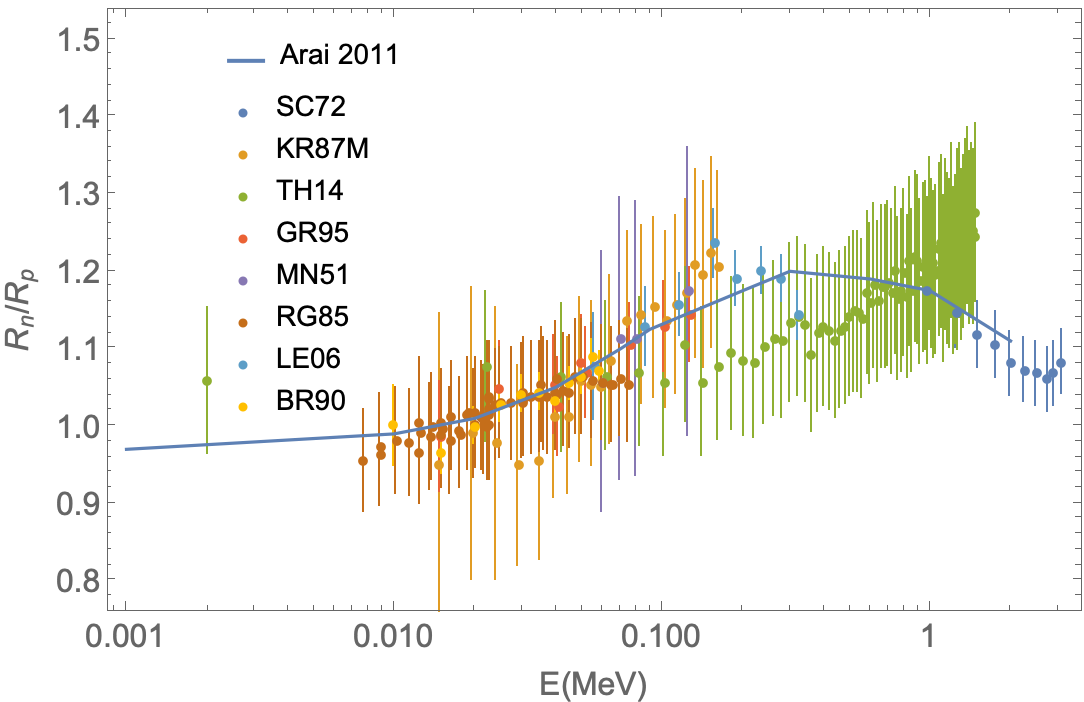}\hspace{2pc}%
\begin{minipage}[b]{.44\textwidth}\caption{\label{ratio_n_su_p}(Color online) Ratios of ddn to ddp S-factors from experiments (dots) and theory \cite{Arai:2011zz} (solid line). The Trojan Horse points from \cite{Tumino:2014}  with their 1-$\sigma$ errors are shown in green (TH).}
\end{minipage}
\end{figure}

Since ddn and ddp are strong interactions with relatively low Coulomb barrier and both have application in the nuclear energy field, the amount of experimental data is sufficiently large if compared with other processes. We use all data considered in our previous analysis \cite{Pisanti:2007hk} adding to them the TH data \cite{Aurora} (all data are reported in Fig. \ref{dd_data}). We checked the level of the discrepancy of these data with theoretical nuclear calculations \cite{Arai:2011zz} pointed out in \cite{Coc:2015bhi} by computing, after the minimization procedure that provides the normalization constants $\omega_{ddn}$ and $\omega_{ddp}$, the uncertainties on the values of the ratio $(\omega_{ddn}^{(TH)}S_{ddn})/(\omega_{ddp}^{(TH)}S_{ddp})$. The TH points from \cite{Tumino:2014} (green dots) with their 1-$\sigma$ errors are shown in Fig. \ref{ratio_n_su_p}, from which it seems that, even if the central values do not follow the theoretical curve, yet they are fully consistent with it within 1-$\sigma$, though it is evident that the energy trend is quite different. In any case, inclusion of TH data have a minor impact on the global data fit and on the rates.

\begin{figure}[p]
\begin{minipage}{.45\textwidth}
\includegraphics[width=1.\textwidth]{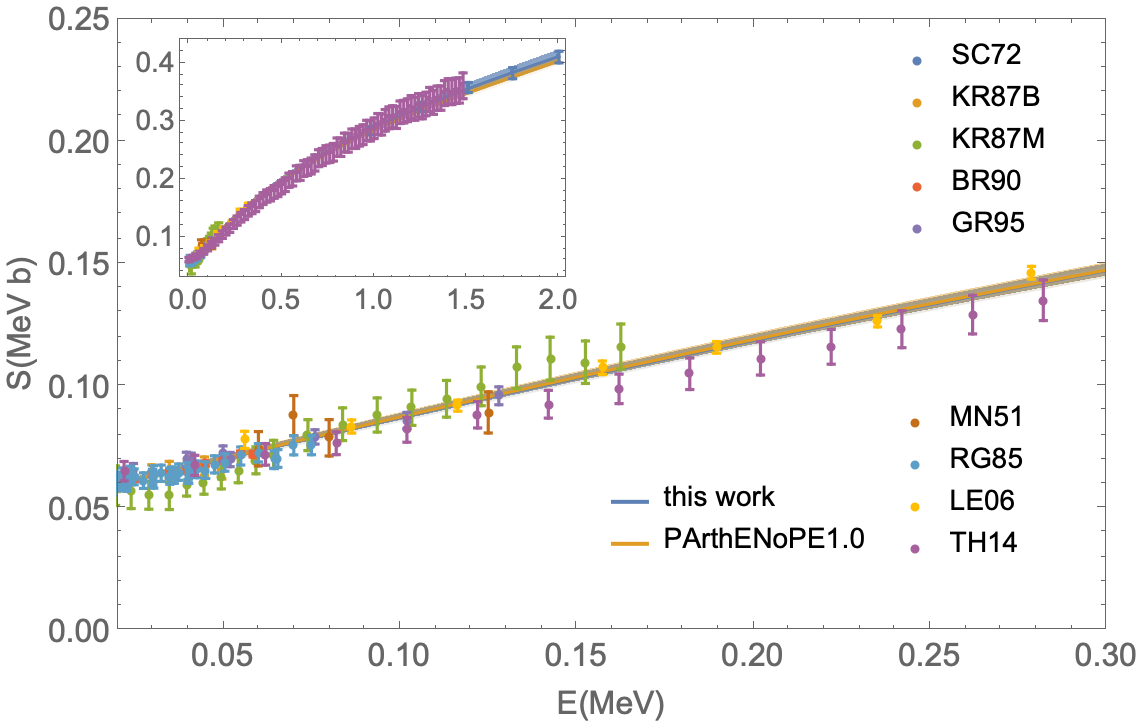}
\end{minipage}\hspace{2pc}%
\begin{minipage}{.45\textwidth}
\includegraphics[width=1.\textwidth]{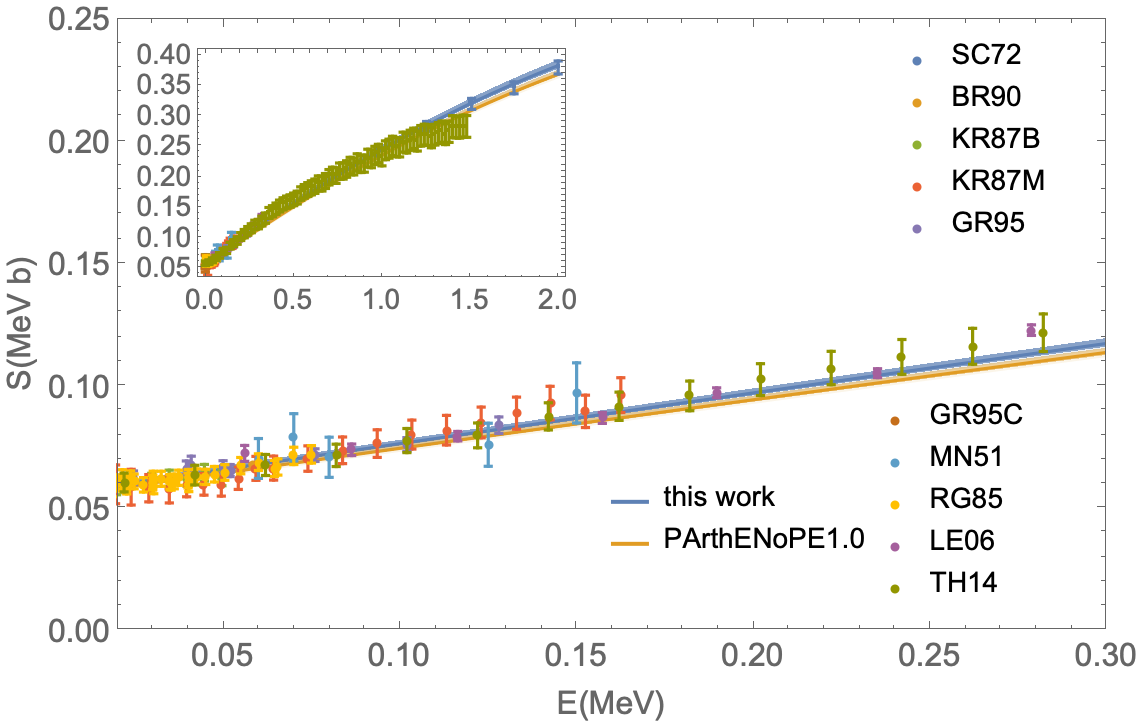}
\end{minipage} 
\caption{\label{S_dd}(Color online) The d-d transfer reactions S-factors versus center of mass energy: ddn (left plot) and ddp (right plot). Note that the experimental points have been renormalized by the corresponding $\omega_k$.  The curves are the data fit derived in this work (blue) and the fit of \cite{Pisanti:2007hk} (orange). Bands are the 1-$\sigma$ uncertainty ranges.}
\end{figure}
\begin{figure}[p]
\begin{minipage}{.45\textwidth}
\includegraphics[width=1.\textwidth]{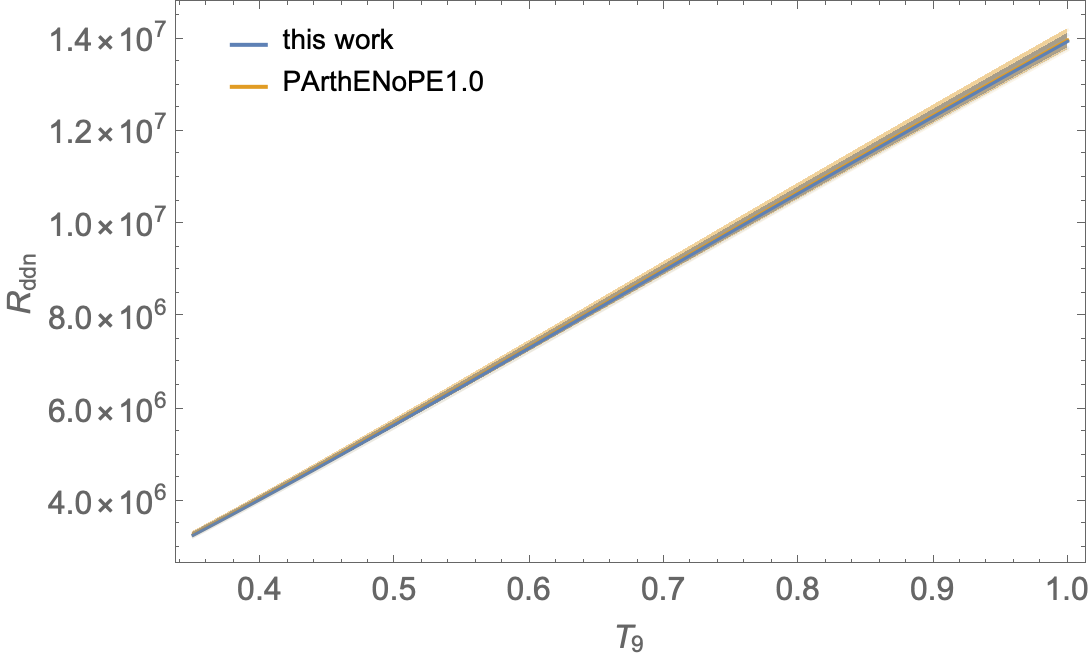}
\end{minipage}\hspace{2pc}%
\begin{minipage}{.45\textwidth}
\includegraphics[width=1.\textwidth]{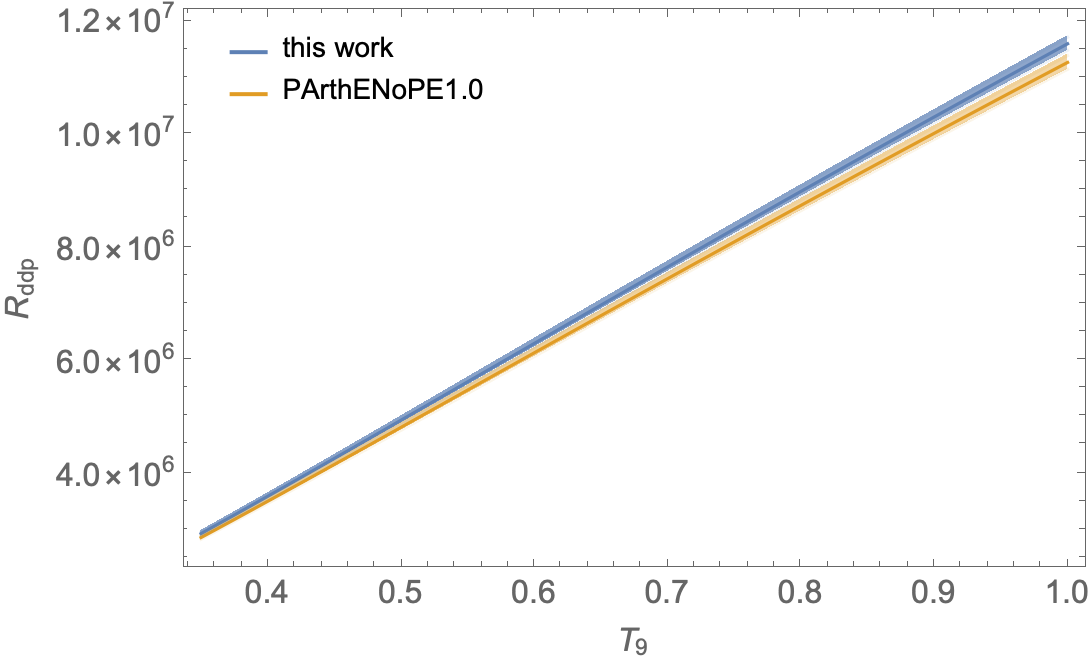}
\end{minipage} 
\caption{\label{R_dd}(Color online) $R_{ddn}$ and $R_{ddp}$ rates versus the temperature in $10^9$ K. Conventions are the same of Fig. \ref{S_dd}.}
\end{figure}
\begin{figure}[p]
\begin{minipage}{.45\textwidth}
\includegraphics[width=1.\textwidth]{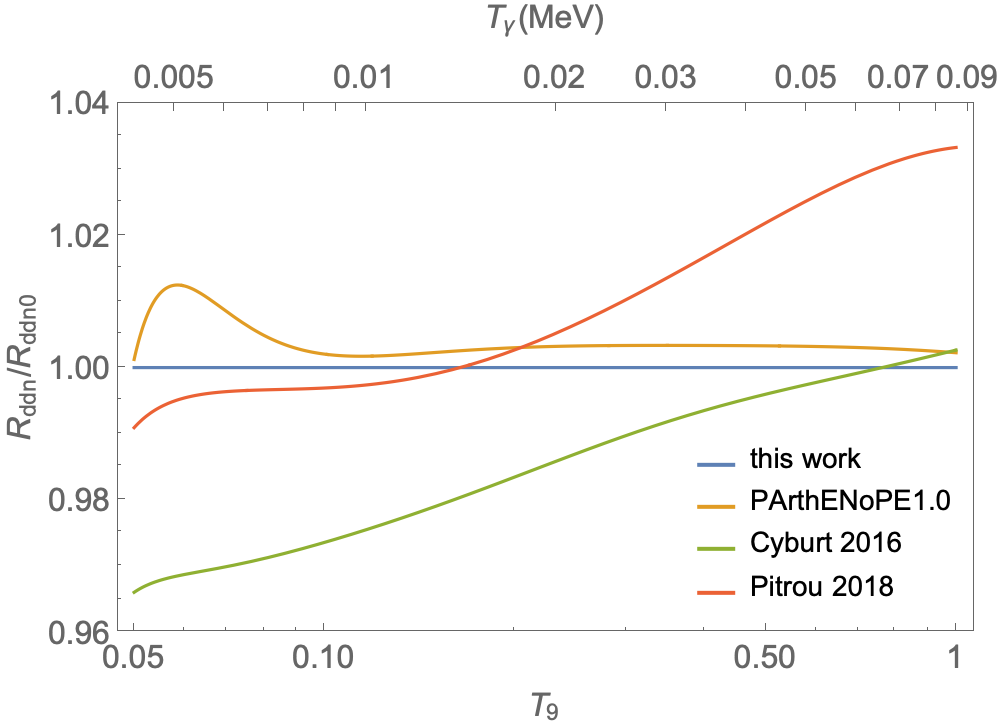}
\end{minipage}\hspace{2pc}%
\begin{minipage}{.45\textwidth}
\includegraphics[width=1.\textwidth]{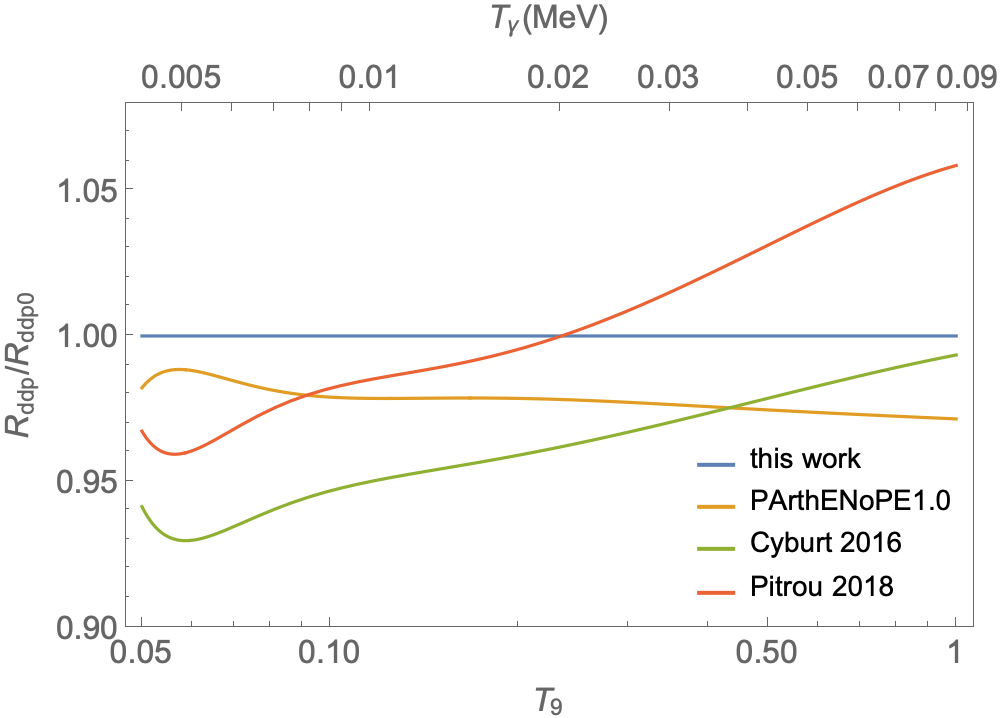}
\end{minipage} 
\caption{\label{dd}Ratios between different $R_{ddn}$ (left plot) and $R_{ddp}$ (right plot) rates and their benchmark values obtained in this work, $R_{ddn0}$ and $R_{ddp0}$, as a function of the temperature in $10^9$ K (see text).}
\end{figure}

In Fig. \ref{S_dd} we plot the results of the fits of the S-factors together with experimental data, while the corresponding rates with their 1-$\sigma$ uncertainty as function of the temperature $T_9\equiv T/(10^9\,\text{K})$ are shown in Fig. \ref{R_dd}. The S-factors together with the normalization factors $\omega_k$ for each data set are reported in \ref{A1}.

Finally, we show in Fig.~\ref{dd} the comparison of our present calculation for d-d with \cite{Pisanti:2007hk} (\paenpe{1.0}), \cite{Cyburt:2015mya} (Cyburt 2016), and \cite{Pitrou:2018cgg} (Pitrou 2018). From the plots it appears that a 3\% (7\%) maximum difference in $R_{ddn}$ ($R_{ddp}$) can be produced by the different analysis methods and data selection, while of course the difference with \paenpe{1.0} is due to the new data included in the analysis.

\subsection{The role of datasets and of statistical analysis for ddn and ddp}

\begin{figure}[t]
\begin{minipage}{.45\textwidth}
\includegraphics[width=1.\textwidth]{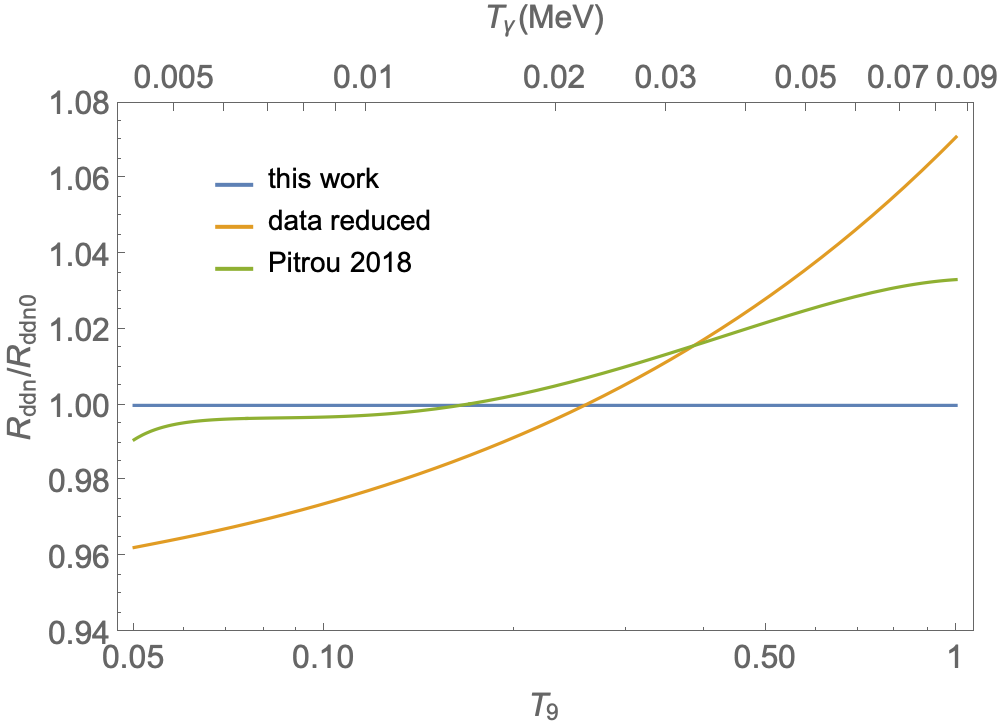}
\end{minipage}\hspace{2pc}%
\begin{minipage}{.45\textwidth}
\includegraphics[width=1.\textwidth]{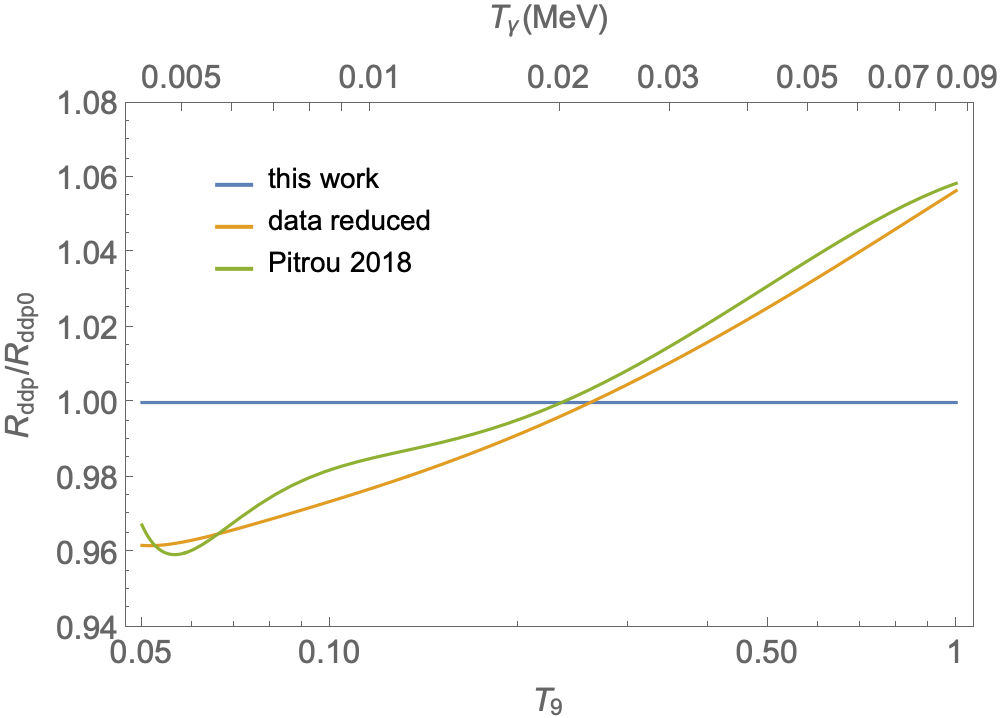}
\end{minipage} 
\caption{\label{comparison} Ratios between the ddn (left) and ddp(right)  rate obtained adopting different datasets and statistical approach, see text, and their benchmark values obtained in this work,  $R_{ddn0}$ and $R_{ddp0}$, versus temperature in $10^9$ K.}
\end{figure}
In order to assess and quantify the effect of the different approaches in nuclear rate evaluation of ddp and ddn, in particular to compare our results with those of \cite{Pitrou:2018cgg}, we repeated our analysis with a different set of experimental data on the S-factor, while still using the $\chi^2$ of Eq. (\ref{chi2tot}). Our results are shown in Fig. \ref{comparison}. In this plot we show the rates $R_{ddn}$ and $R_{ddp}$, normalized to the determination of these rates presented in this paper ($R_{ddn0}$ and $R_{ddp0}$), in two cases. In the first case we performed a polynomial fit of the S-factors by considering the data used in \cite{Pitrou:2018cgg} only (``data reduced", orange curves). In the second case we show the rates found in \cite{Pitrou:2018cgg} (``Pitrou 2018", green lines).  

Consider first the ddp case (right plot). We see that by selecting data as \cite{Pitrou:2018cgg} we obtain very similar results to \cite{Pitrou:2018cgg}, compare the ``data reduced" and ``Pitrou 2018" curves,  showing that the main source of difference between our findings and \cite{Pitrou:2018cgg} is due to the data selection criterium. The residual difference in the BBN range can be attributed  to the different expression of the $\chi^2$ adopted, see (\ref{chi2tot}) and (\ref{chi2pitrou}), and possibly,  to the different function used in the fit of the S-factor, a polynomial fit in our case, a nuclear theory motivated (smooth) functional form in \cite{Pitrou:2018cgg}. The role of the two different fitting function seems however, subleading since once the same data are used, the temperature behaviour of the rates is very close, and they differ by approximately a constant factor over the relevant temperature range. The ddn case (left plot) is quite different. Even adopting the same datasets, our result in this case (orange curve) is quite different than the result of \cite{Pitrou:2018cgg}, showing that, {\it for the reduced set of data} the assumed fitting function plays a major role. While excluding data at higher energy, say outside the BBN range, affects quite strongly the polynomial fit in the ddn case, compare the orange curve with the benchmark result described in this paper $R_{ddn0}$ (the blue horizontal line), adopting a theoretical motivated fitting function gives a more stable result when some datasets are not considered in the analysis, see the green line. Yet, the difference found in this case with respect to our estimate $R_{ddn0}$ shows the impact of choosing a reduced sample of data. As we stated already, if there are no clear indication of large systematics in a particular experiment, we don't think it is worth neglecting any of available data, even those outside the BBN energy range. 

\subsection{Primordial Deuterium sensitivity to $S(E)$ factors}
\label{sensitivity}

\begin{figure}[b]
\includegraphics[width=.5\textwidth]{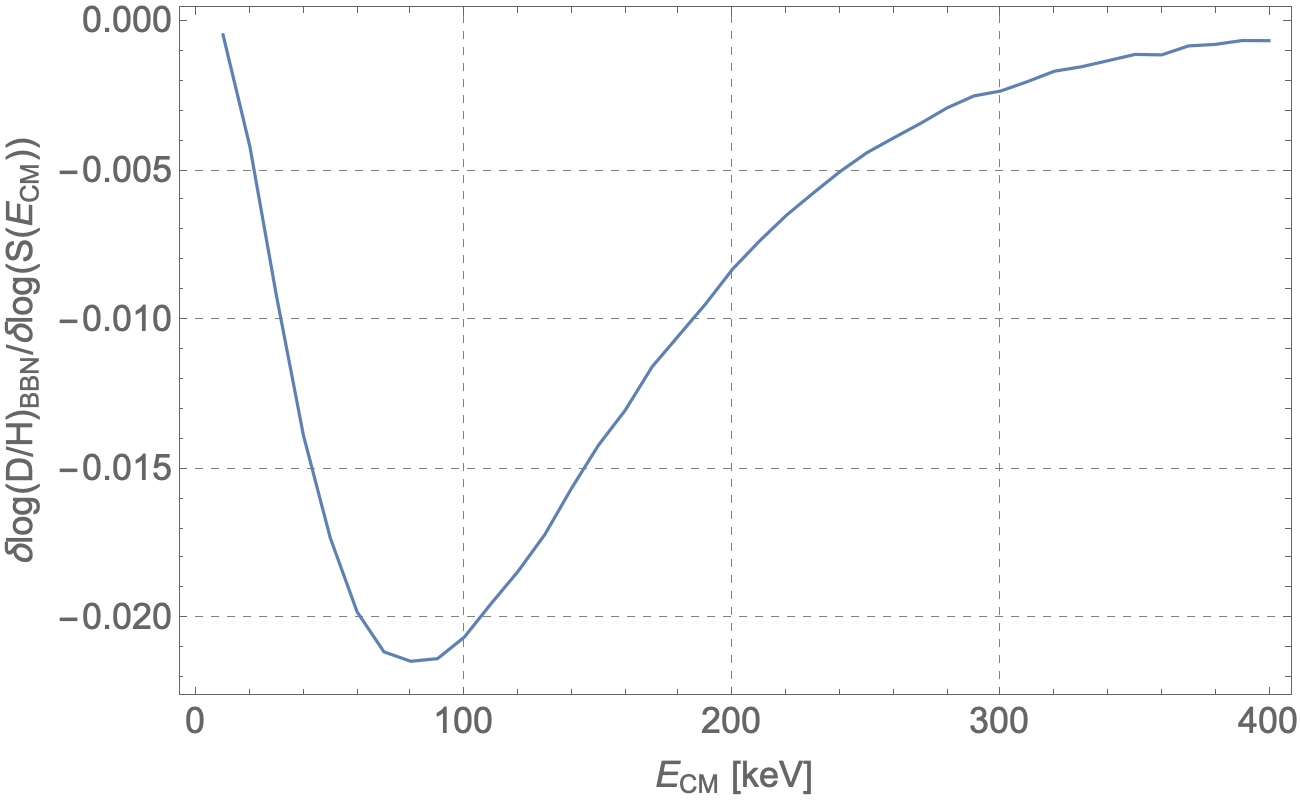}\hspace{2pc}%
\begin{minipage}[b]{.44\textwidth}\caption{\label{sensitivity1} The primordial deuterium sensitivity to dp$\gamma$ S-factor versus the center of mass energy $E_{CM}$.}
\end{minipage}
\end{figure}
Deuterium synthesis and burning takes place in a temperature range of the nucleon-photon plasma roughly of the order of $T \sim 100 - 20$ keV. This is the temperature interval where one expects that precise measurements of Deuterium  production and burning rates can have a big impact in improving the accuracy of the theoretical prediction of its primordial yield. From Eq. (\ref{rate}) we see that this range translates into an analogous one for the center of mass energy  where the three main process S-factors considered in this paper require a measurement as precise as possible. A quantitative way to determine more accurately the relevant center of mass energy range of a given reaction process for deuterium prediction is obtained in terms of the {\it sensitivity function}, see e.g. \cite{Fiorentini:1998fv,Nollett:2000fh}, defined as the logarithmic derivative of the D/H abundance versus the corresponding $S$ factor
\begin{equation}
\sigma(E_{\rm{CM}})= \frac{\delta {\rm D/H}}{{\rm D/H}}\frac{S(E_{\rm{CM}})}{\delta S(E_{\rm{CM}})}.
\end{equation}
\begin{figure}[t]
\begin{minipage}{.45\textwidth}
\includegraphics[width=1.\textwidth]{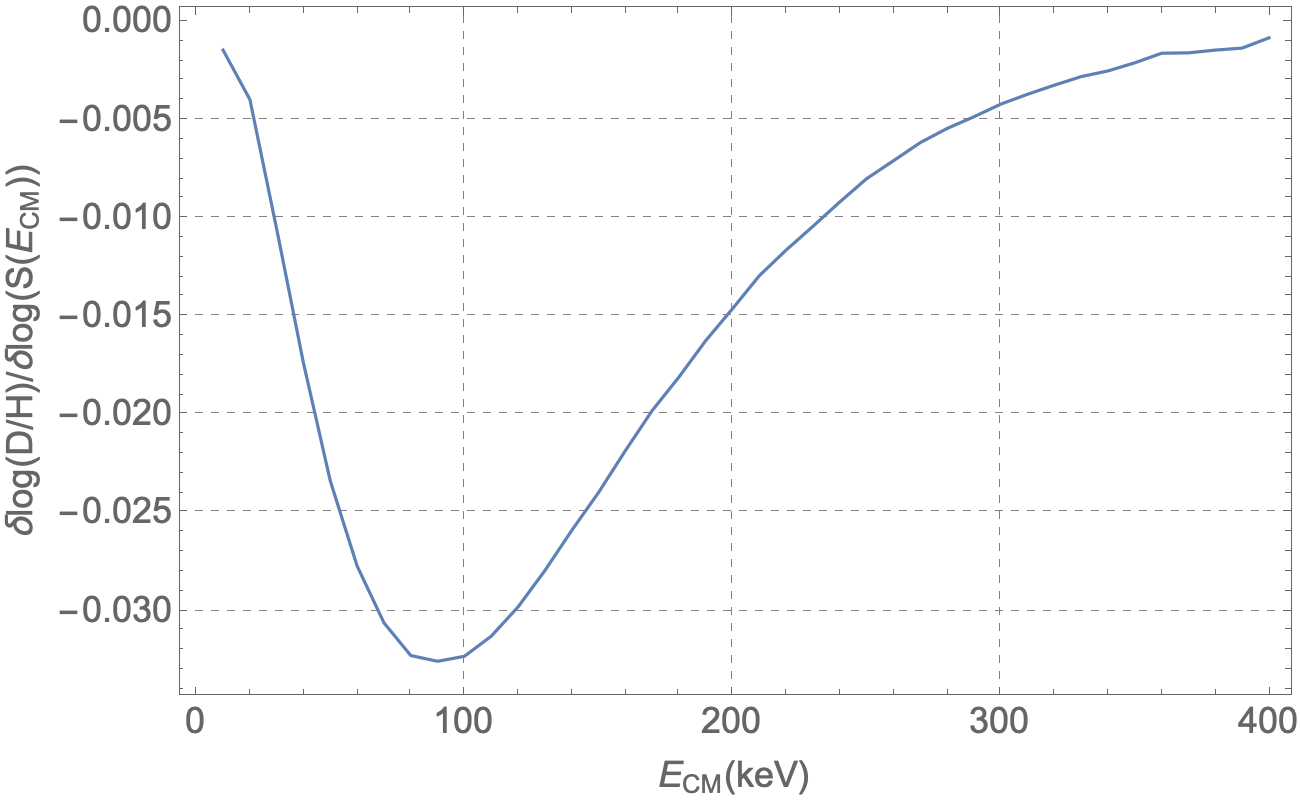}
\end{minipage}\hspace{2pc}%
\begin{minipage}{.45\textwidth}
\includegraphics[width=1.\textwidth]{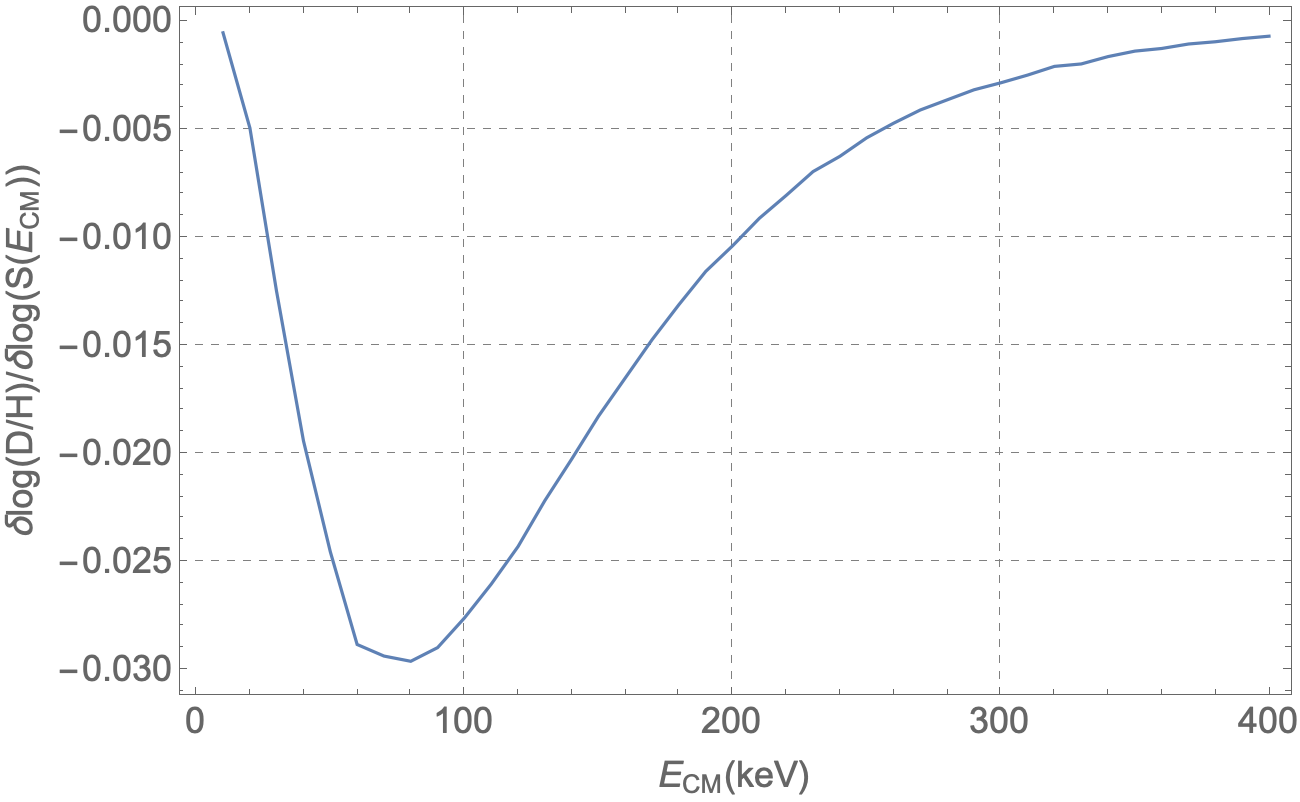}
\end{minipage} 
\caption{\label{sensitivity2}The primordial deuterium sensitivity to ddn (left) and ddp (right) $S$ factor versus the center of mass energy $E_{CM}$.}
\end{figure}
We have computed  $\sigma(E_{\rm{CM}})$ by varying the $S$ factor for all energy bins, with a spacing of 10 keV, in the whole range $10-500$ keV, and computing the corresponding rate by a convolution with the thermal nucleon distribution as function of the energy. The corresponding yield of deuterium is obtained using the {\paenpe} code. The results are shown in Fig.s \ref{sensitivity1} and \ref{sensitivity2}. We see that the D abundance is sensitive to the dp$\gamma$, ddn and ddp cross sections mainly in the range $E_{\rm CM} =20-240$ keV with the largest effect at $E_{\rm} \sim 80$ keV. Notice why the new LUNA measurement of the dp$\gamma$ S-factor are so important, since they cover the whole sensitivity region.

\section{Update of \paenpe}
\label{paenpe}

The new determinations of $R_{dp\gamma}$, $R_{ddn}$, and $R_{ddp}$ rates and their 1-$\sigma$ uncertainties, considered in the previous section, have been implemented in the public code \paenpe~\cite{Gariazzo:2019}, while the rate for the p(n,$\gamma$)$^2$H, $R_{pn\gamma}$, is unchanged. To have an understanding of the role of each rate in determining the Deuterium abundance one can consider the propagated uncertantainty on D/H when that rate is changed in its 1-$\sigma$ range {\it keeping all others fixed}. These are reported in Table \ref{D_errors}. From this table it is evident the decreased contribution to the error budget of D(p,$\gamma)^3$He uncertainty (column two) with respect to the previous situation \cite{DiValentino:2014cta} (column four), due to the very precise determination from LUNA experiment. This Table is only for illustrative purpose. Of course, a reliable estimate of the total error on Deuterium prediction is obtained by randomly varying all rates simultaneously in their 1-$\sigma$ range, and in fact,  it turns to be larger than if we sum the results of Table \ref{D_errors}. In fact, we have taken this approach to estimate the D/H theoretical uncertainty and the result is reported later.
\begin{table}[t]
\begin{center}
\begin{tabular}{ccccc}
\br
& $\sigma_{\text D}^{(i)}\cdot 10^5$ & $\delta\sigma_i^2/\sigma_{\rm tot}^2$ (\%) & $(\sigma_{\text D }^{(i)})_{\rm old} \cdot 10^5$ & $(\delta\sigma_i^2/\sigma_{\rm tot}^2)_{\rm old}$ (\%)\\
\mr
$R_{pn\gamma}$ & 0.002 & 0.3 & 0.002 & 0.1 \\
$R_{dp\gamma}$ & 0.027 & 58.5 & 0.062 & 87.0 \\
$R_{ddn}$ & 0.018 & 26.9 & 0.020 & 9.1 \\
$R_{ddp}$ & 0.013 & 14.2 & 0.013 & 3.8 \\
\br
\end{tabular}
\end{center}
\caption{\label{D_errors}Contributions to the theoretical Deuterium uncertainty from the nuclear rates and their relative weight: $\sigma_{\text D}^{(i)}$ represents the uncertainty on Deuterium abundance due to 1-$\sigma$ variation of the corresponding rate in the first column. Notice that these results refer to the case when a single rate is changed at a time. The label $old$ refers to the corresponding previous results obtained in \cite{DiValentino:2014cta}.}
\end{table}

In the standard scenario the only free parameter entering the BBN dynamics is the value of the baryon to photon number density, $\eta$. We remind that the relation between $\eta_{10} \equiv \eta \times 10^{10}$ and $\omega_b$ weakly depends on the $^4$He mass fraction, $Y_p$,
\be
\omega_b=\frac{1-0.007125\,Y_p}{273.279} \left ( \frac{G_N}{6.70728\times 10^{-45}\, {\rm MeV}^{-2}} \right ) \left ( \frac{T^0_{\gamma}}{2.7255\, {\rm K}} \right )^3 \eta_{10},
\ee
with $G_N$ the Newton constant and $T^0_{\gamma}$ the CMB temperature today.
In non standard scenarios, another input to BBN calculations is the extra relativistic degrees of freedom, also called number of extra effective neutrino species,
\be
\dN =\neff - 3.045,
\ee
related to possible additional components to the radiation energy density at the BBN epoch, in addition to photons and to the three active neutrinos (and/or to a neutrino distributions in phase space with some nontrivial features, such as chemical potentials). The standard model expectation for $\neff$, 3.045 \cite{deSalas:2016ztq} instead of 3, is related to several corrections in the treatment of neutrino decoupling, among which radiative corrections to the relevant quantities (see also \cite{Mangano:2005cc} and references therein). We report in \ref{A3} our fits of D/H ratio and Helium mass fraction $Y_p$ as function of $\omega_b$ and $\dN$. The fit accuracy is better than 0.08\% for Helium and 0.1\% for Deuterium on the ranges $0.01\leq \omega_b \leq 0.03$ and $-3\leq \dN \leq 3$.

We also updated the scaling relations \cite{Cyburt:2004cq} of Deuterium and Helium abundances with respect to variations of model parameters,
\be
Y_i=Y_{i,0}\prod_n\left ( \frac{p_n}{p_{n,0}} \right )^{\alpha _n},\,\,\, Y_i= Y_p, \mbox{D/H}.
\ee 
The $p_n$ represent the various input quantities ($\omega_b$, $\tau_n$ and $\neff$) to the BBN calculations as well as key nuclear rates which affect the abundance $Y_i$ while $p_{n,0}$ are their fiducial input values. The exponents $\alpha_n$ are the sensitivities, defined as the logarithmic derivatives of the light-element abundances with respect to each variation of our fiducial parameters \cite{Fiorentini:1998fv}. These scalings, normalized at $\omega_b = 0.02242 \pm 0.00014$, $\tau_n = (879.4 \pm 0.6)$ and $\neff=3.045$, are:
\be
\frac{D}{\text{H}}=2.51\times10^{-5}\, R_{pn\gamma}^{-0.20}\, R_{dp\gamma}^{-0.31}\, R_{ddn}^{-0.51}\, R_{ddp}^{-0.42}\left ( \frac{\omega_b}{0.02242} \right )^{-1.61}\left ( \frac{\tau_n}{879.4\, {\rm s}} \right )^{0.43} \left(1+ \frac{\Delta \neff}{3.045} \right)^{0.41},
\label{scalingD}
\ee
\be
Y_p=0.2469\, R_{pn\gamma}^{0.005}\, R_{dp\gamma}^{0.0002}\, R_{ddn}^{0.006}\, R_{ddp}^{0.005}\left ( \frac{\omega_b}{0.02242} \right )^{0.04}\left ( \frac{\tau_n}{879.4\, {\rm s}} \right )^{0.72}\left(1+ \frac{\Delta \neff}{3.045} \right)^{0.16},
\label{scalingHe}
\ee
\\
with no substantial variation, as expected, with respect to those found in \cite{Serpico:2004gx}. Of course, the scaling behaviour is quite accurate only in a neighborhood
of the chosen fiducial values for the various parameters, yet large enough to cover their present uncertainty range.

\begin{table}[t]
\caption{\label{deuterium}Values of (D/H)$\cdot 10^5$ obtained by different groups by varying the rate used for $R_{dp\gamma}$. Results are shown for $\omega_b = 0.02225$.}
\begin{center}
\begin{tabular}{cccc}
\br
$R_{dp\gamma}$ & \texttt{PArthENoPE2.1} & Pitrou {\it et al.} (2018) \cite{Pitrou:2018cgg} & Cyburt {\it et al.} (2016) \cite{Cyburt:2015mya} \\
\mr
this work     & $2.54\pm 0.07$ & & \\
MAR2005 \cite{Marcucci:2005zc}     & $2.52\pm 0.07$ & $2.459\pm 0.036$ & \\
AD2011 \cite{Adelberger:2010qa} & $2.58\pm 0.07$ & & 2.579 \\
MAR2015 \cite{Marcucci:2015yla}   & $2.45\pm 0.07$ & & \\
\br
\end{tabular}
\end{center}
\end{table}
In order to make the comparison with previous analyses, we here adopt $\omega_b = 0.02225\pm 0.00016$ (as in table II of \cite{Cyburt:2015mya}) and a standard number of relativistic degrees of freedom $\neff=3.045$ during BBN \cite{Mangano:2005cc,deSalas:2016ztq}. Table \ref{deuterium} compares our results with those of Pitrou {\it et al.} (2018) \cite{Pitrou:2018cgg} and Cyburt {\it et al.} (2016) \cite{Cyburt:2015mya}, for different choice of the rate $R_{dp\gamma}$. While there is a nice agreement with Deuterium obtained in \cite{Cyburt:2015mya} when the corresponding $R_{dp\gamma}$ rate (AD2011) is used in \paenpe, a 2.4\% difference is found between our finding with $R_{dp\gamma}$ given by MAR2005 and the one in \cite{Pitrou:2018cgg}, which can be attributed to the different approaches that we have described in fitting the ddn and ddp rates.

\section{Cosmological implications}
\label{cosmology}

We now discuss the implications of our results on the determination of cosmological parameters. At this aim, we compare our theoretical results with the Deuterium astrophysical measurement of \cite{Cooke:2017cwo}
\be
^2\text{H/H}=(2.527 \pm 0.030)\times 10^{-5}.
\ee
For the $^4$He mass fraction we will consider the four determinations already mentioned in the Introduction \cite{Peimbert:2016bdg,Aver:2015iza,Izotov:2014fga,Hsyu:2020uqb},
\bea
Y_p&=& 0.2446\pm 0.0029,\\
Y_p&=& 0.2449\pm 0.0040,\\
Y_p&=& 0.2551\pm 0.0022,\\
Y_p&=& 0.2436 \pm 0.0040.
\eea
We use the latest Planck result for the baryon density \cite{Planck2018} with (Planck+BAO) and without (Planck) combination with BAO data, see Section 1, Eq.s \ref{Planck} and \ref{Planck+BAO}.

\subsection{Likelihood analysis}

\begin{figure}[b]
\includegraphics[width=.5\textwidth]{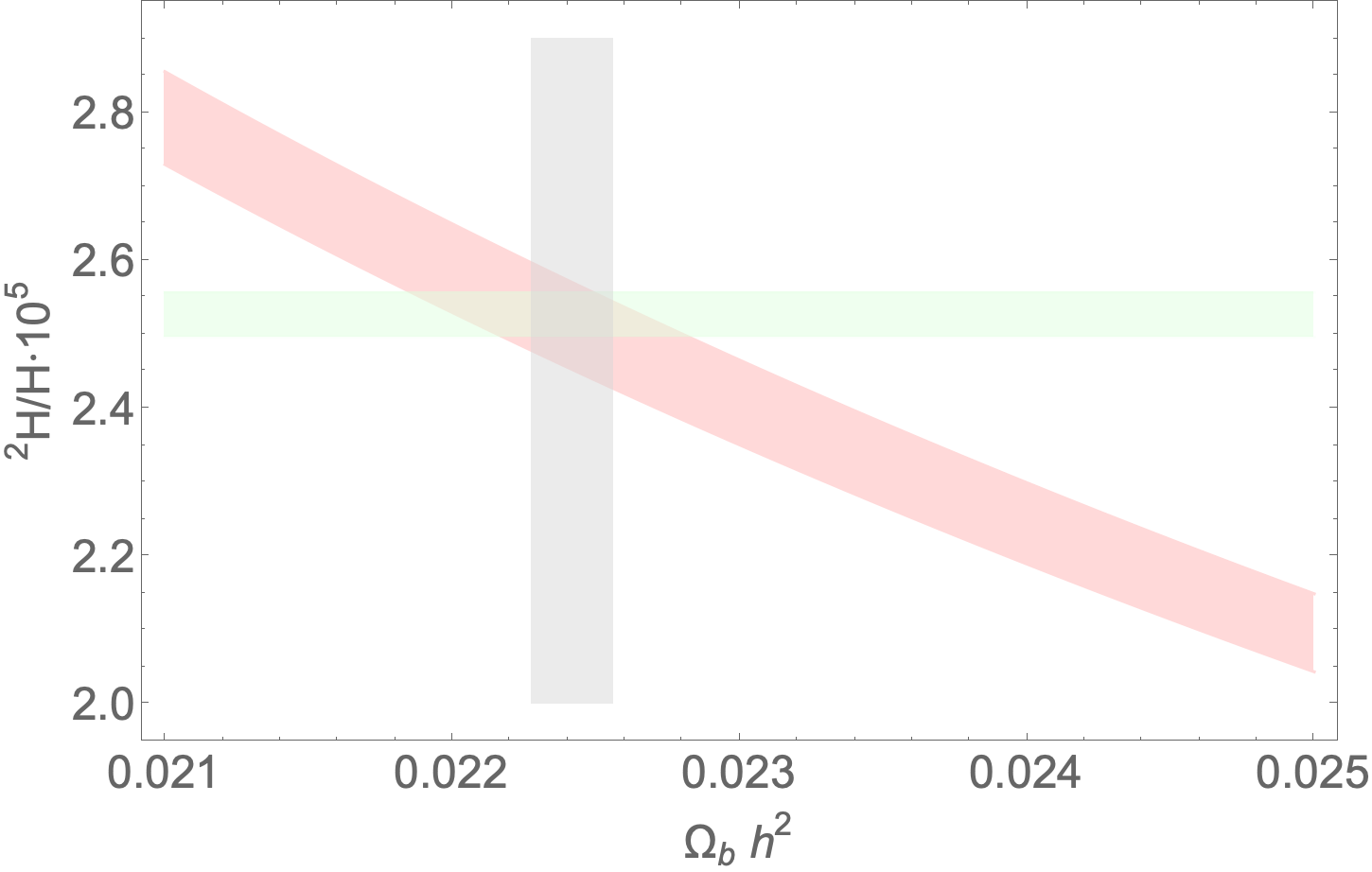}\hspace{2pc}%
\begin{minipage}[b]{.44\textwidth}\caption{\label{D_evolution}Values of the theoretical primordial Deuterium abundance with 68\%\,C.L.\,(red band) as a function of $\omega_b$, calculated for $\dN=0$. The green horizontal band represent the experimental determination \cite{Cooke:2017cwo} with 1-$\sigma$ error while the gray vertical band is the CMB measurement by Planck at 1-$\sigma$.}
\end{minipage}
\end{figure}

The predicted abundance of Deuterium obtained in the standard (minimal) scenario with $\dN = 0$, when $\omega_b$  is fixed by the Planck determination, is D/H = $\left( 2.51 \pm 0.06\pm 0.03\right) \times 10^{-5}$, where the two errors are obtained by propagation of the uncertainties on nuclear rates and baryon density, respectively, in nice agreement with the measured one (see Fig.\,\ref{D_evolution}). To get this error all dp$\gamma$, ddn and ddp rates have been all changed randomly within their 1-$\sigma$ range. Once the overall consistency of the model has been checked, it is interesting to investigate how Deuterium abundance can constrain the baryon density in the Standard BBN scenario with three standard active neutrinos and $\omega_b$ the only free parameter, {\it without} using the Planck result. We use a standard likelihood analysis where the one-dimensional likelihood function
\be
\mathcal{L}_{\mbox{\small D}-3\nu}(\omega_b) = \mathcal{L}_D(\omega_b,0),
\label{LSBBN}
\ee
with
\be
\mathcal{L}_D(\omega_b, \neff) \propto \text{exp}\left ( -\frac{\chi_D^2(\omega_b, \neff)}{2} \right ),
\label{LSBBN}
\ee
and
\be
\chi_D^2(\omega_b, \neff)= \frac{(Y_D^{\text{(th)}}\,(\omega_b, \neff)-\,Y_D^{\text{(exp)}})^2}{\sigma_D^{\text{(th)}2}(\omega_b, \neff)+\sigma_D^{\text{(exp)}2}},
\ee
with $\sigma_D^{\text{(th)}}(\omega_b, \neff)$ and $\sigma_D^{\text{(exp)}}$ the theoretical and experimental uncertainties on Deuterium, respectively. We name this standard scenario as  D-$3\nu$. 
\begin{figure}[t]
\begin{center}
\includegraphics[width=.5\textwidth]{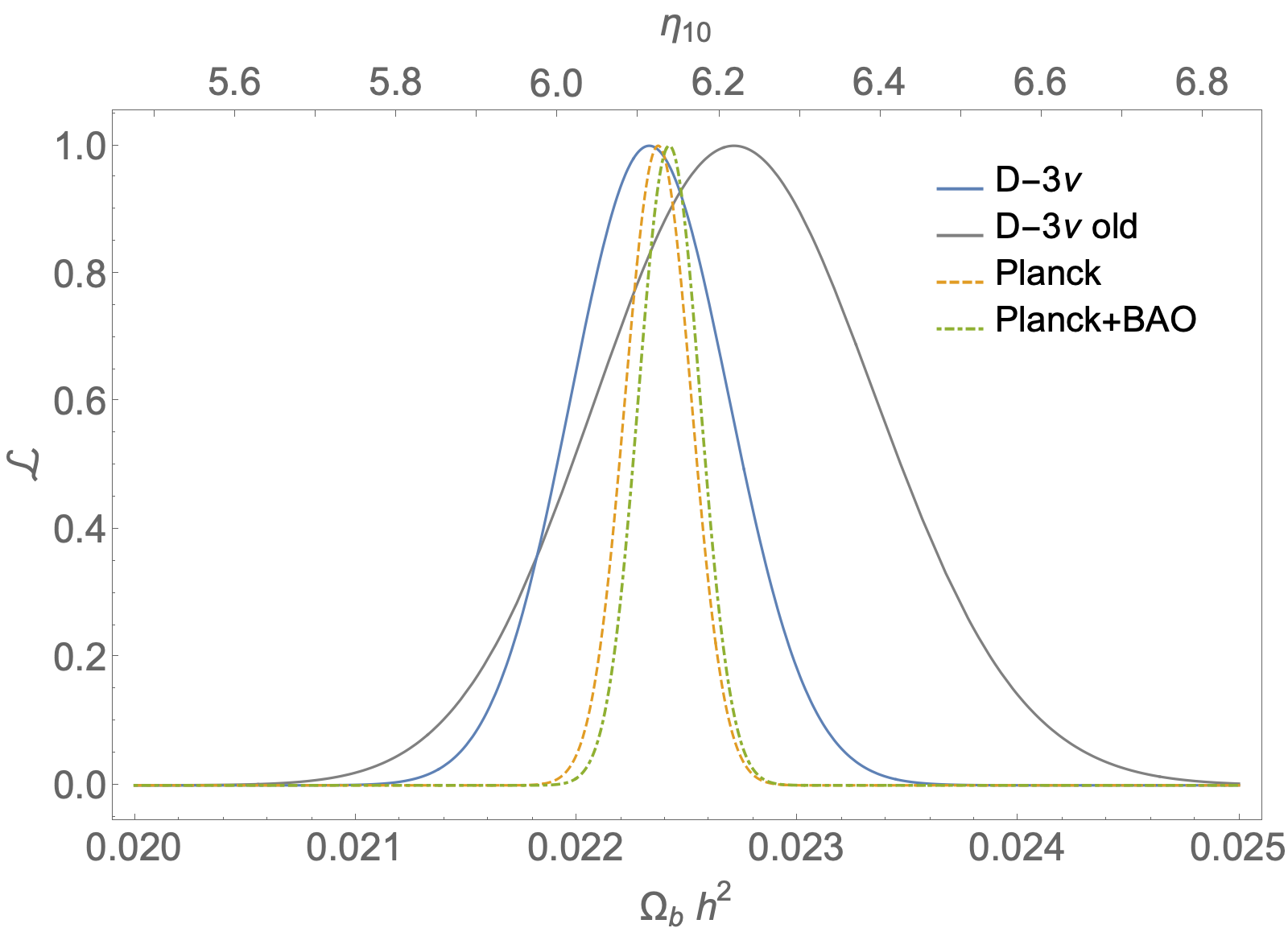}\hspace{2pc}
\end{center}
\begin{minipage}[t]{\textwidth}\caption{\label{like1}One-dimensional likelihoods for a standard BBN scenario ($\dN$=0) as a function of $\omega_b$ using Deuterium abundance (D-3$\nu$) (solid blue line) compared with the Planck 2018 TT, TE, EE, lowE and lensing results combined with (green dashed line) and without (orange dashed line) BAO data. The broader grey solid line is the old determination from Deuterium before LUNA data on the dp$\gamma$ cross section but using our new determination of the ddn and ddp rates.}
\end{minipage}
\end{figure}
Fig. \ref{like1} shows the impressive agreement of $\mathcal{L}_{\mbox{\small D}-3\nu}$ (solid blue curve) with Planck results (Planck: dashed orange curve; Planck +BAO: green dashed curve). By marginalizing $\mathcal{L}_{\mbox{ \small D}-3\nu}$ we obtain $\omega_b = 0.02233 \pm 0.00036$ at 68\% C.L., which shows that the use of BBN Deuterium alone as a baryometer is reaching a precision which is presently only a factor 2 worst than what can be obtained by  CMB results of Planck experiment, thanks to the remarkable improvements on astrophysical D measurements and the new LUNA results on the dp$\gamma$ rate, compare the blue (including LUNA data) and grey (without LUNA data) curves.

\begin{figure}[t]
\begin{center}
\begin{tabular}{cc}
\begin{minipage}{.3\textwidth}
\includegraphics[width=1.\textwidth]{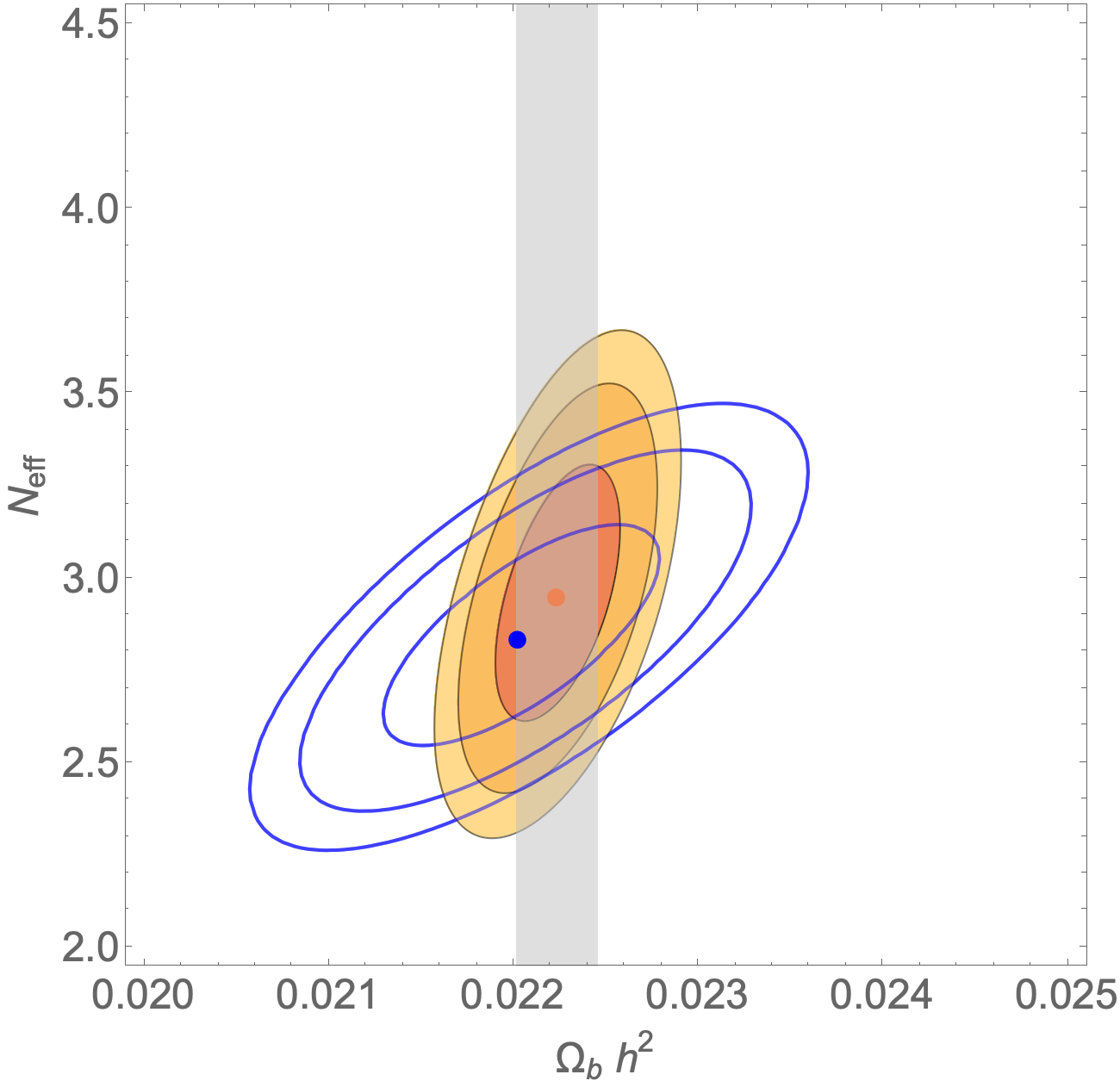}
\end{minipage}\hspace{1pc}&
\begin{minipage}{.3\textwidth}
\includegraphics[width=1.\textwidth]{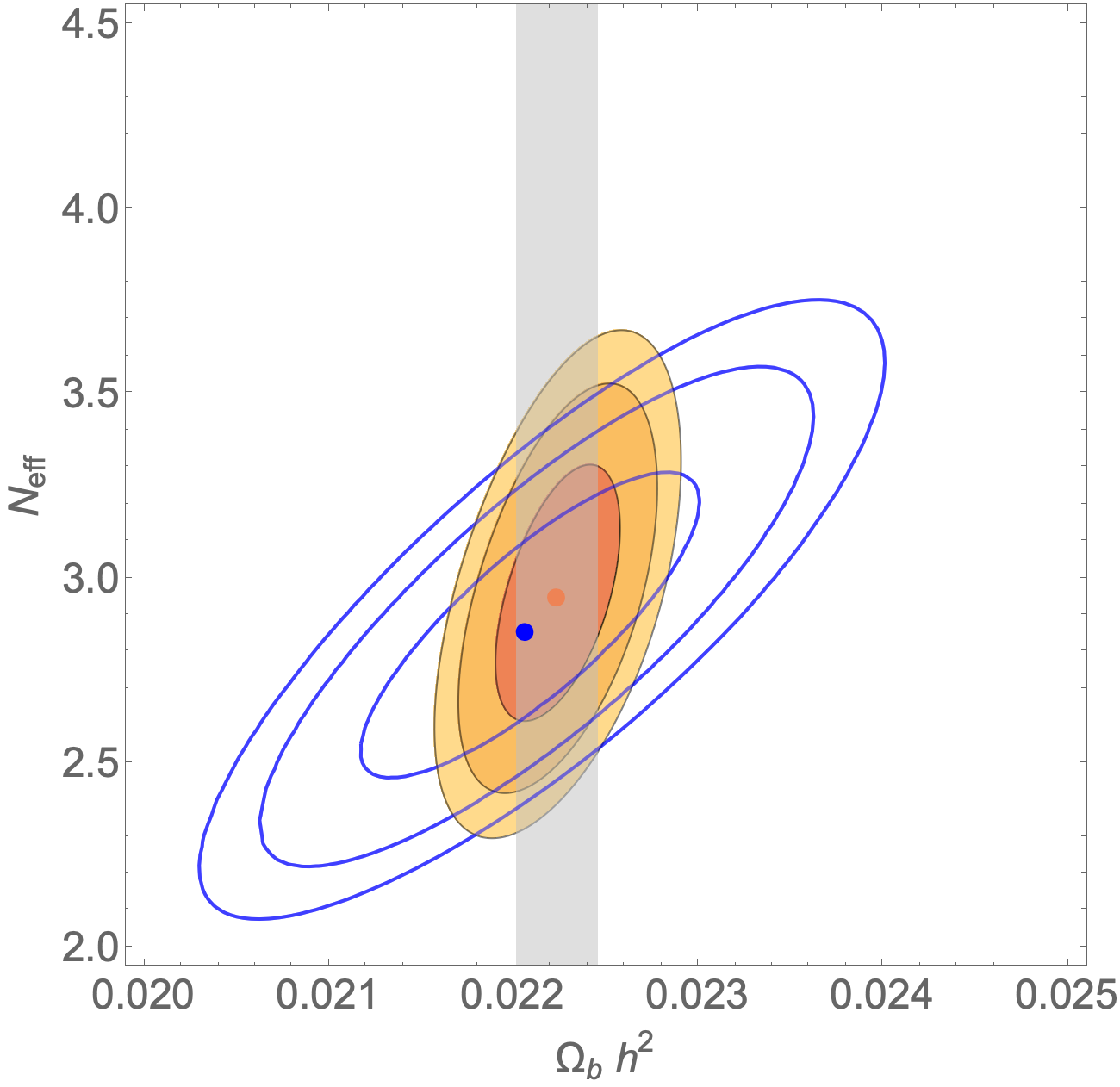}
\end{minipage} \hspace{1pc}\\
\begin{minipage}{.3\textwidth}
\includegraphics[width=1.\textwidth]{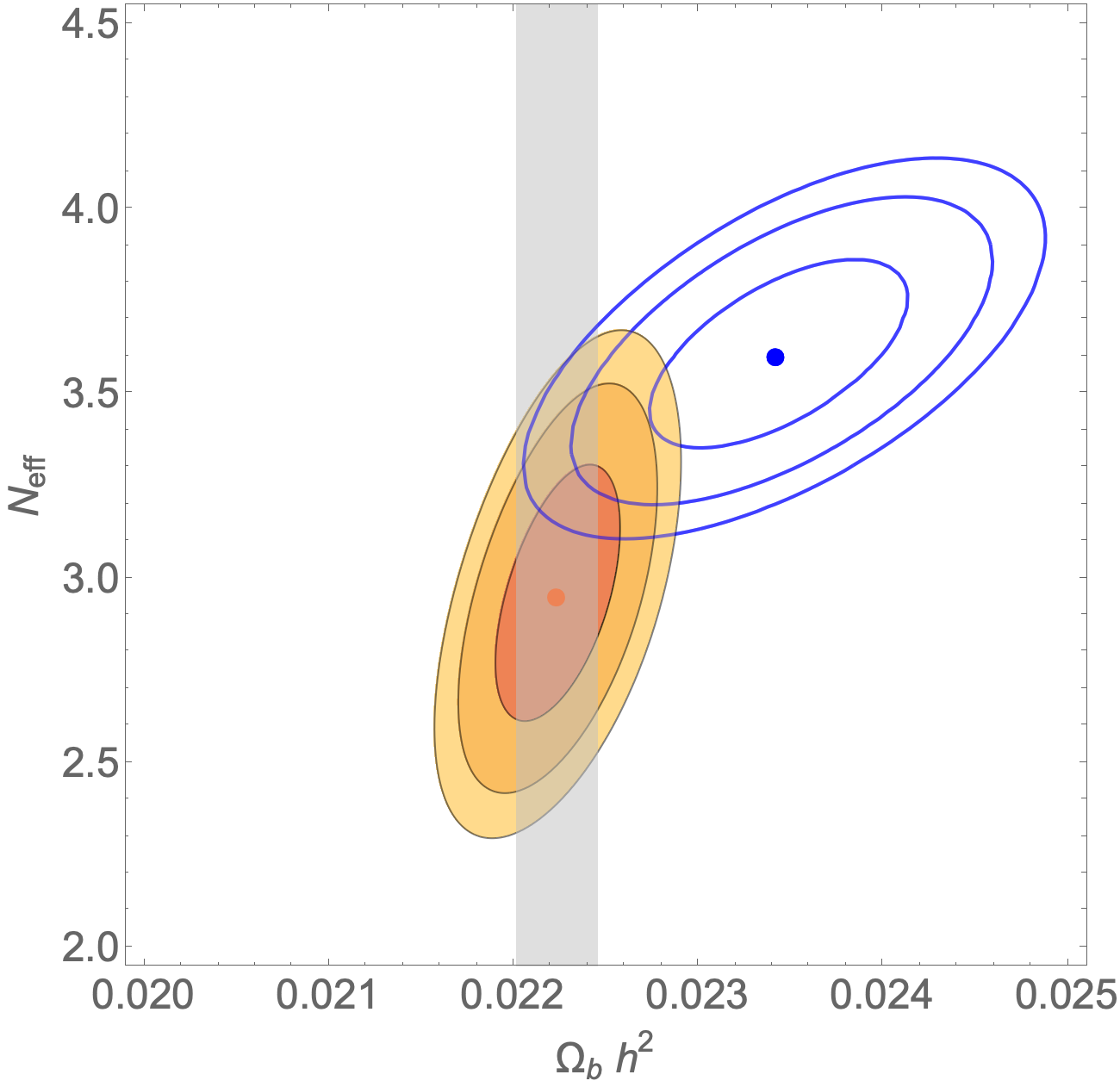}
\end{minipage}&
\begin{minipage}{.3\textwidth}
\includegraphics[width=1.\textwidth]{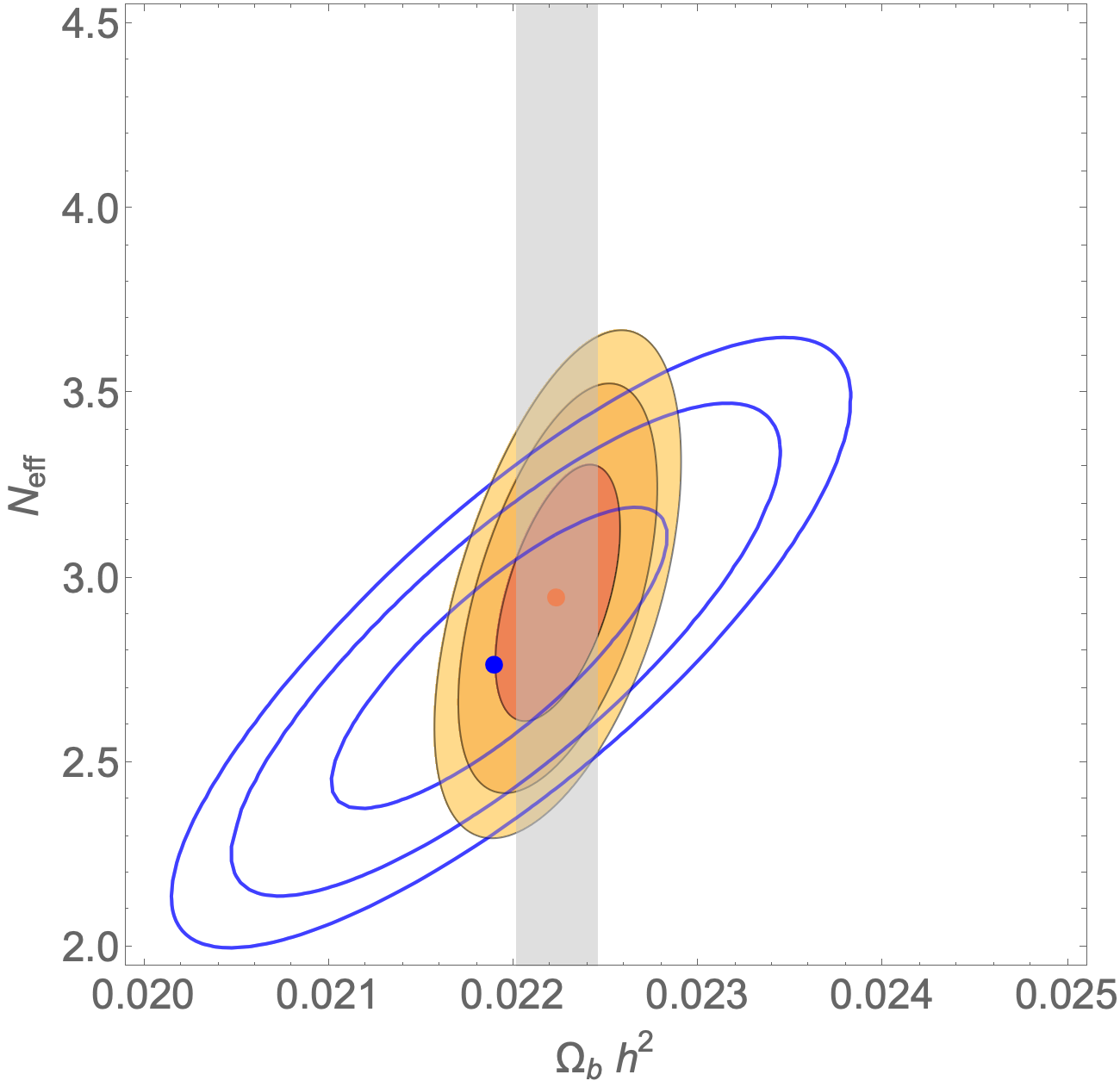}
\end{minipage} 
\end{tabular}
\end{center}
\caption{\label{likelihoods}Likelihood contours (68, 95 and 99\% C.L.) in the $\omega_b$-$\neff$ plane using different $Y_p$ estimates: top left \cite{Peimbert:2016bdg}, top right \cite{Aver:2015iza}, bottom left \cite{Izotov:2014fga}, bottom right \cite{Hsyu:2020uqb}. The smaller filled contours corresponds to $\mathcal{L}_{\rm D+Planck}(\omega_b,\neff)$, while the broader solid blue ones are obtained using $\mathcal{L}_{\rm BBN}(\omega_b,\neff)$. The vertical band is the Planck 2018 determination of $\omega_b$ at 68\%C.L., with a free $\neff$.}
\end{figure}
We now consider  a cosmological model where $\neff$ is a free parameter. In this case, the Deuterium likelihood alone is unable to break the degeneracy between $\omega_b$ and $\neff$. It is possible to break this degeneracy in two ways, the first one by using a gaussian prior on the baryon density as measured by Planck when $\neff$ is considered as a free parameter, namely $\omega_{\text{Planck}} = 0.02224 \pm 0.00022$,
\be
\mathcal{L}_{\text{Planck}}(\omega_b) \propto \text{exp}\left(-\frac{(\omega_b - \omega_{\text{Planck}})^2}{2\, \sigma_{\text{Planck}}^2}\right).
\label{prior}
\ee
so that the likelihood function is now chosen as
\be
\mathcal{L}_{\rm D+Planck}(\omega_b,\neff) = \mathcal{L}_D(\omega_b,\neff)\cdot \mathcal{L}_{\rm Planck}(\omega_b).
\label{LD_Planck}
\ee
Alternatively one can also exploit the measured value of primordial $^4$He, i.e. multiplying that Deuterium likelihood by a similar term for $Y_p$
\be
\mathcal{L}_{\rm BBN}(\omega_b,\neff) = \mathcal{L}_D(\omega_b,\neff)\cdot \mathcal{L}_{\rm He}(\omega_b,\neff).
\label{LBBN}
\ee

Note that the predicted $^4$He mass fraction in  standard BBN when $\omega_b = \omega_{\text{Planck}}$ is $Y_p = 0.2469\pm 0.0001$, which is compatible in 1-$\sigma$ with the measurements of \cite{Peimbert:2016bdg},\cite{Aver:2015iza} and \cite{Hsyu:2020uqb}, while is slightly more than 3-$\sigma$ away from the results of \cite{Izotov:2014fga}. Because of the  large experimental error on $Y_p$ we expect that using the $^4$He mass fraction $prior$ will give a looser constraint.

We show in Fig. \ref{likelihoods} the likelihood contours (68, 95 and 99\% C.L.) in the $\omega_b$-$\neff$ plane. The smaller orange filled contours corresponds to $\mathcal{L}_{\rm D+Planck}(\omega_b,\neff)$, i.e. the Deuterium abundance with the Planck prior on $\omega_b$, while the broader solid blue ones are obtained using both Deuterium and $Y_b$ and no prior on the baryon density, i.e. using $\mathcal{L}_{\rm BBN}(\omega_b,\neff)$. The four different plots correspond to the determination of $Y_p$  of \cite{Peimbert:2016bdg},\cite{Aver:2015iza}, \cite{Izotov:2014fga} and \cite{Hsyu:2020uqb}. The vertical band is the Planck 2018 determination of $\omega_b$ at 68\%C.L. in case of a free $\neff$. By marginalizing the D+Planck likelihood over $\omega_b$ or $\neff$ we obtain $\neff = 2.95 \pm 0.22$ and $\omega_b = 0.02224 \pm 0.00022$, respectively, while in the BBN case we get $\neff = 2.84\pm 0.20$ and $\omega_b = 0.0220 \pm 0.0005$ when using \cite{Peimbert:2016bdg}, $\neff = 2.86 \pm 0.28$ and $\omega_b = 0.0221 \pm 0.0006$ adopting the results of \cite{Aver:2015iza}, and finally, $\neff = 2.78 \pm 0.28$ and $\omega_b = 0.0219 \pm 0.0006$ for the value of $Y_p$ found in \cite{Hsyu:2020uqb}, all compatible with a standard value of $\neff$. On the other hand, the $^4$He mass fraction found by \cite{Izotov:2014fga} suggest a larger radiation content in the BBN epoch, at 3-$\sigma$ level, namely $\neff = 3.60+0.17$ and a large baryon density  $\omega_b = 0.0234 \pm 0.0005$. In this latter case the standard BBN scenario may be reconsidered, by introducing a larger number of relativistic species which is found to be larger than its standard value at the 3-$\sigma$ level. The discrepancy on the baryon density is milder but we note that {\it both} parameters are somehow at variance with the present standard cosmological scenarios, and may possibly require both some dark radiation {\it and} a mechanism to explain the different value of $\omega_b$ at BBN and CMB epochs. It is quite evident however, that it would be extremely important to first understand the role of systematics in the astrophysical measurement of $Y_p$ and the disagreement among different results before abandoning the Ockham razor.

\begin{table}[t]
\begin{center}
\begin{tabular}{ccc}
\br
& $\omega_b$ & $\neff$ \\
\mr
{\bf Planck} & $0.02237\pm 0.00015$ & 3.045 \\
{\bf Planck+BAO} & $0.02242\pm 0.00014$ & 3.045 \\
{\bf D-$3\nu$} & $0.02233\pm 0.00036$ & 3.045 \\
{\bf D+Planck} & $0.02224 \pm 0.00022$ & $2.95 \pm 0.22$ \\
{\bf BBN} \cite{Peimbert:2016bdg} & $0.0220 \pm 0.0005$ & $2.84\pm 0.20$ \\
{\bf BBN} \cite{Aver:2015iza} & $0.0221 \pm 0.0006$ & $2.86 \pm 0.28$ \\
{\bf BBN} \cite{Izotov:2014fga} & $0.0234 \pm 0.0005$ & $3.60 \pm 0.17$ \\
{\bf BBN} \cite{Hsyu:2020uqb} & $ 0.0219 \pm 0.0006 $ & $ 2.78 \pm 0.28$ \\
\br
\end{tabular}
\end{center}
\caption{\label{marginal}Summary of the results of different analysis described in the text. Uncertainties are at 68\% C.L..}
\end{table}
Table \ref{marginal} summarizes our findings for $\omega_b$ and $\neff$ obtained in the various cases by marginalizing the corresponding likelihoods, together with the CMB results from Planck experiment.

\section{Conclusions}
\label{conclusions}

The recent data obtained by the LUNA collaboration for the D(p,$\gamma$)$^3$He $S$ factor represent a remarkable achievement and  experimental challenge, reaching an uncertainty of 3\% \cite{Mossa:2020qgj,Nature}. Its impact on the theory and predictions of BBN is comparable to that of the exquisite measurements of primordial Deuterium in Quasar Absorption Systems, now at the level of 1\% \cite{Cooke:2017cwo}. The combination of these two sets of data have one major consequence: Deuterium produced during BBN is presently an excellent  probe for baryon density and fundamental physics. In particular, as shown in \cite{Nature}, the value of $\omega_b$ can be fixed by the D/H ratio with an increased accuracy, which is not far from what has been obtained by the Planck experiment on CMB anisotropies, actually only a factor 2.4 larger. Improvements in Deuterium measurements foreseen in the near future and ab initio theoretical nuclear calculations of the most relevant few body processes, see e.g. \cite{Marcucci:2015yla,Marcucci:2005zc} for the dp$\gamma$ process,  may further refine this result.
Furthermore, the new best fit value for the S-factor measured at LUNA, higher than previous determinations, gives a lower Deuterium abundance, in agreement with astrophysical determination, and pins down a value of $\omega_b$ in very good agreement with the Planck result, confirming at a high precision level, the robustness of the standard cosmological model.

In this paper we have described a new analysis of the three main processes responsible for Deuterium burning during BBN, the dp$\gamma$ and the two exchange reactions ddp and ddn, describing the statistical method used in obtaining the corresponding thermal rate from experimental data, the data selection criteria, and the resulting theoretical prediction for the D/H ratio, the latter obtained using an updated version of the public PArthENoPE code. The new LUNA data and its impact on the thermal rate for dp$\gamma$ and Deuterium abundance has been already presented in \cite{Nature}. Yet, we have also included here its discussion, describing in more details the analysis and more technical aspects for completeness.  

Actually, a fraction of order 40\% of the total error budget on D/H is represented by ddn and ddp reactions. A new analysis of the impact of these reactions is one of the main issues discussed in this paper.   We have reported a detailed study of all present data on $R_{dp\gamma} $ and $R_{ddp}$ rates, which has been also used to obtain the cosmological results described in \cite{Nature}, and compared our results with other similar studies, which differ for data selection and statistical method in data analysis. We found that both these aspects lead to, at worst, a less than 2-$\sigma$ difference among our results and what has been found elsewhere, see e.g. \cite{Pitrou:2018cgg}. We think that a closer collaboration among different groups in assessing a common understanding and quantifying the theoretical primordial D/H uncertainty would be quite important and desirable. 

As for the cosmological implications, we have already stressed the agreement of the baryon density found using primordial Deuterium theoretical prediction and its astrophysical measurements with the Planck result, and its improved precision due to LUNA results.

We have also discussed in this paper the case where  the relativistic species energy density, $\neff$, is left as a free parameter, and how the different measurements of the primordial $^4$He mass fraction $Y_p$ affects the conclusions on how large is the amount of dark radiation compatible with present data. We remind that the latest Planck results give $\neff=2.99 \pm 0.17$ at 68\% C.L. \cite{Planck2018}. When $\neff$ is not fixed, because of the degeneracy between $\neff$ and $\omega_b$, we have combined the D theoretical prediction with either a Planck prior on baryon density or using only BBN data, i.e. the $^4$He mass fraction in addition to deuterium and no prior on $\omega_b$. In the first case  we find $\neff = 2.95 \pm 0.22$ at 68\% C.L..
On the other hand, the second case is still controversial, since there are two groups of measurements of primordial $Y_p$ which (in the best case) disagree at the 2.5-$\sigma$ level. On one side, the results of \cite{Peimbert:2016bdg,Aver:2015iza,Hsyu:2020uqb} combined with Deuterium abundance data and measurements, give values of $\omega_b$ and $\neff$ in fair agreement with the concordance model and with the standard theoretical expectation $\neff=3.046$. Differently, the higher value found in \cite{Izotov:2014fga} pin down a 3-$\sigma$ evidence for dark radiation, in addition to the standard contribution of the three active neutrino species, as well as a mild disagreement of the baryon density with respect to the Planck results.
It is clear that better measurement of pristine abundance of $^4$He and a reduction of systematic effects on $Y_p$ would be of great impact in the future in this respect.

\ack
Gianpiero Mangano and Ofelia Pisanti warmly thank the members of the LUNA Collaboration. Discussions and sharing of expertises have been a great experience. Work supported by the Italian grant 2017W4HA7S ``NAT-NET: Neutrino and Astroparticle Theory Network'' (PRIN 2017) funded by the Italian Ministero dell'Istruzione, dell'Universit\`a e della Ricerca (MIUR), and Iniziativa Specifica TAsP of INFN.

\appendix
\section{S-factors}
\label{A1}
In this section we report our polynomial fits of the S-factor for \mbox{D(p, $\gamma)^3$He} radiative capture and the deuteron-deuteron transfer reactions, D(d, n)$^3$He and D(d, p)$^3$H, the covariance matrix between the coefficients and the normalization factors for each data set determined using the MINUIT package \cite{minuit}. Note that the small differences with respect to the expression of the S-factor reported in \cite{Nature} are due to the inclusion of data of \cite{Tisma:2019acf}.

\begin{itemize}

\item
\mbox{D(p, $\gamma)^3$He}
\be
S(E) = (0.2121+5.975\,E+5.463\,E^2-1.665\,E^3)\cdot 10^{-6}\,\text{MeV\,b}, \nn
\ee

\be
\text{cov}(a_i,a_j)=10^{-15}\cdot\left(
\begin{array}{cccc}
0.0140 & -0.378 & 1.07  & -0.462  \\
-0.378  & 29.5    & -90.0 & 39.5  \\
1.07     & -90.0   & 479.  & -230. \\
-0.462  & 39.5    & -230. & 112. \\
\end{array}
\right), \nn
\ee
\medskip

\be
\begin{array}{lll}
\omega_{\rm LU02}=0.9950, & \omega_{\rm SC72} = 1.051, & \omega_{\rm GR63} = 0.9872, \\
\omega_{\rm GR62} = 1.025, & \omega_{\rm MA97} = 1.073, & \omega_{\rm WA63} = 1.022, \\
\omega_{\rm GE67} = 1.001, & \omega_{\rm LU20} = 0.9978, & \omega_{\rm TI19} = 0.9819.
\end{array} \nn
\ee
\medskip

\item
D(d, n)$^3$He
\be
S(E) = (0.05225 + 0.3655\, E - 0.1799\, E^2 + 0.05832\, E^3 - 0.007393\, E^4)\, \text{MeV\,b}, \nn
\ee

\be
\text{cov}(a_i,a_j)=10^{-6}\cdot\left(
\begin{array}{ccccc}
0.0699 & -0.600 & 0.893 & -0.454 & 0.0738 \\
-0.600  & 16.0    & -26.2  & 13.9    & -2.30    \\
0.893   & -26.2   & 54.0   & -30.6   & 5.29     \\
-0.454  & 13.9    & -30.6  & 18.1    & -3.22    \\
0.0738 & -2.30   & 5.29   & -3.22   & 0.586   \\
\end{array}
\right), \nn
\ee
\medskip

\be
\begin{array}{lll}
\omega_{\rm SC72}=1.019, & \omega_{\rm KR87B} = 1.105, & \omega_{\rm KR87M} = 1.017, \\
\omega_{\rm BR90} = 0.9986, & \omega_{\rm GR95} = 0.9547, & \omega_{\rm MN51} = 1.133, \\
\omega_{\rm RG85} = 0.9653, & \omega_{\rm LE06} = 0.9977, & \omega_{\rm TU14} = 1.029.
\end{array} \nn
\ee
\medskip

\item
D(d, p)$^3$H
\be
S(E) = (0.05520+0.2151\,E-0.02555\,E^2)\, \text{MeV\,b}, \nn
\ee

\be
\text{cov}(a_i,a_j)=10^{-6}\cdot\left(
\begin{array}{ccc}
0.0384 & -0.117 & 0.0243 \\
-0.117  & 2.38    & -0.413  \\
0.0243 & -0.413 & 0.0936 \\
\end{array}
\right), \nn
\ee
\medskip

\be
\begin{array}{lll}
\omega_{\rm SC72}=1.039, & \omega_{\rm BR90} = 1.009, & \omega_{\rm KR87B} = 1.110, \\
\omega_{\rm KR87M} = 1.054, & \omega_{\rm GR95} = 0.9702, & \omega_{\rm GR95C} = 0.9977, \\
\omega_{\rm MN51} = 1.019, & \omega_{\rm RG85} = 0.9771, & \omega_{\rm LE06} = 0.9880, \\
\omega_{\rm TU14} = 1.008.
\end{array} \nn
\ee

\end{itemize}

\newpage 

\section{Deuterium and Helium abundances}
\label{A3}
In this section we report our fits of the main nuclide abundances as function of $\omega_b$ and $\dN$. The fitting functions we employ are
\be
\sum_n \sum_m a_{nm}\, \omega_b^{n-1}\, \dN^{m-1}, \nn
\ee
and the coefficients are reported below. The fit accuracy is better than 0.08\% for Helium and than 0.1\% for Deuterium on the ranges $0.01\leq \omega_b \leq 0.03$ and $-3\leq \dN \leq 3$.
\begin{itemize}

\item
D/H

\be
a = \left(
\begin{array}{cccc}
7.4930781\cdot10^{-4} & 1.1653472\cdot10^{-4} & 6.290673\cdot10^{-6} & -7.4481692\cdot10^{-7} \\
-0.17962147 & -0.029246304 & -2.0550163\cdot10^{-3} & 3.4355779\cdot10^{-4} \\
21.264426 & 3.5653547 & 0.29138904 & -0.063198045 \\
-1485.3582 & -255.00941 & -23.025325 & 6.1447324 \\
64085.229 & 11250.861 & 1086.7839 & -344.1985 \\
-1.6828042\cdot10^6 & -302010.18 & -30515.713 & 11173.375 \\
2.4702764\cdot10^7 & 4.5304844\cdot10^6 & 471076.15 & -195500.17 \\
-1.5564167\cdot10^8 & -2.9152507\cdot10^7 & -3.0818069\cdot10^6 & 1.4274025\cdot10^6
\end{array}
\right), \nn
\ee
\bigskip

\item
$Y_p$

\be
a = \left(
\begin{array}{cccc}
0.219017 & 8.95576\cdot10^{-3} & -5.16856\cdot10^{-3} & 3.96732\cdot10^{-4} \\
2.59343 & 1.55837 & 1.58478 & -0.110667 \\
-5.61747 & -237.504 & -254.63 & 16.9548 \\
-12989. & 20072.2 & 22201.4 & -1375.53 \\
1.00162\cdot10^6 & -1.00573\cdot10^6 & -1.13728\cdot10^6 & 63629.8 \\
-3.54918\cdot10^7 & 2.97276\cdot10^7 & 3.42807\cdot10^7 & -1.66343\cdot10^6 \\
6.28119\cdot10^8 & -4.78915\cdot10^8 & -5.63747\cdot10^8 & 2.23671\cdot10^7 \\
-4.48551\cdot10^9 & 3.24118\cdot10^9 & 3.90654\cdot10^9 & -1.15002\cdot10^8
\end{array}
\right), \nn
\ee

\end{itemize}

\end{document}